\documentclass[showpacs,aps,prd,reprint,superscriptaddress,nofootinbib,longbibliography]{revtex4-1}
\usepackage[colorlinks=true, pdfstartview=FitV, linkcolor=magenta,citecolor=blue, urlcolor=blue,
bookmarks=true, bookmarksnumbered=true, breaklinks]{hyperref}
\usepackage[dvipdfmx]{graphicx}
\usepackage{amsmath,amssymb,bm,color,longtable,mathrsfs,slashed}
\usepackage{ulem}

\newcommand{\Slash}[1]{{\ooalign{\hfil/\hfil\crcr$#1$}}}

\begin{document}

\title{Comprehensive study of mass modifications of light mesons in nuclear matter  in the three-flavor extended Linear Sigma Model}

\author{Daiki Suenaga}
\email[]{\tt suenaga@mail.ccnu.edu.cn}
\affiliation{Key Laboratory of Quark and Lepton Physics (MOE) and Institute of Particle Physics, Central China Normal University, Wuhan 430079, China}

\author{Phillip Lakaschus}
\email[]{\tt lakaschus@th.physik.uni-frankfurt.de}
\affiliation{Institute for theoretical physics, Max-von-Laue Str. 1, D-60438 Frankfurt am Main, Germany}

\date{\today}

\begin{abstract}
We present a comprehensive study of mass modifications of scalar, pseudo-scalar, vector, and axial-vector mesons in nuclear matter using the three-flavor extended Linear Sigma Model (eLSM) and the two-flavor Parity Doublet Model (PDM). The meson masses in nuclear matter are determined by calculating the one-loop nucleon corrections to the meson mean fields. As a result, we find all spin-$0$ meson masses except those of the pion, kaon, and the lightest scalar-isoscalar mesons decrease at finite baryon density. For spin-$1$ mesons, masses of all axial-vector mesons decrease in medium, and the density dependences of the $\rho$ and $\omega$ meson masses strongly depend on the value of chiral invariant mass ($M_0$). Also, our results suggest $M_0\approx0.8\,  {\rm GeV}$ is preferable. 
\end{abstract}

\pacs{}

\maketitle

\section{Introduction}
\label{sec:Introduction}
One of the most important phenomena of Quantum Chromodynamics (QCD) is the spontaneous breakdown of chiral symmetry. This effect is essential to explain the masses of light mesons as well as the interactions among them in the vacuum~\cite{Gasser:1983yg,Gasser:1984gg}. In nuclear matter, however, chiral symmetry is believed to be (partially) restored, which is expected to lead to significant modifications of light meson properties. Therefore, investigating meson properties such as masses in medium gives us clues to a better understanding of the partial (incomplete) restoration of chiral symmetry (see Ref.~\cite{Hatsuda:1994pi,Hayano:2008vn} for reviews and references therein).

While the small masses of the pion and kaon can be well understood by the spontaneous breakdown of chiral symmetry, the large mass of $\eta'$ meson is mainly explained by the $U(1)_A$ axial anomaly effect~\cite{tHooft:1986ooh}. The $U(1)_A$ anomaly is related to the existence of the {\it instanton} which can play an important role in color confinement~\cite{Schafer:1996wv}. Some previous studies suggest the strength of the $U(1)_A$ anomaly can be changed in finite baryon density, but it is still under discussion whether the magnitude of the anomaly is strengthened or weakened~\cite{Bernard:1987sx,Fejos:2016hbp,Fejos:2017kpq}. In association with the change of $U(1)_A$ anomaly in nuclear matter, a mass reduction of the $\eta'$ meson leads to the possibility of the formation of $\eta'$ mesic nuclei as well~\cite{Bass:2005hn,Costa:2002gk,Nagahiro:2006dr,Sakai:2013nba,Sakai:2016vcl,Jido:2018aew,Bass:2018xmz}.

The spectroscopy experiment of the pionic atom at GSI was performed to observe the partial restoration of chiral symmetry in nuclear matter, whose result suggests a reduction of the chiral order parameter: $f_\pi^*(\rho_0)^2/f_\pi^2 \approx 0.64$ at normal nuclear density $\rho_B=\rho_0$~\cite{Suzuki:2002ae}. Also, the fixed-target experiments for vector meson mass modifications in nuclei at J-PARC~\cite{Muto:2005za} and at Jefferson Laboratory~\cite{Wood:2008ee} were operated, but the result is still under discussion due to a complexity by a broadening of vector mesons. Then, another experiment called E16 experiment is planned at J-PARC. Furthermore, the $\eta'$ mesic nuclei experiments at GSI~\cite{Itahashi:2012ut,Tanaka:2016bcp,Tanaka:2017cme} and at the University of Bonn~\cite{Nanova:2012vw,Nanova:2013fxl,Nanova:2016cyn,Friedrich:2016cms,Nanova:2018zlz}, and the $pp\to pp\eta'$ reaction experiment at the COSY accelerator complex~\cite{Czerwinski:2014yot} have been carried out, aimed at the change of $U(1)_A$ axial anomaly in nuclear matter. Another $\eta'$ mesic nuclei experiment is ongoing at SPring-8~\cite{Muramatsu:2013tdv} (see Ref.~\cite{Metag:2017yuh} for a review and references therein).

In the present work, we comprehensively study the mass modifications of light scalar, pseudo-scalar, vector, and axial-vector mesons in nuclear matter to provide useful information on the partial restoration of chiral symmetry and the change of the $U(1)_A$ anomaly in medium to existing and forthcoming experiments. For this purpose, we employ the three-flavor extended Linear Sigma Model (eLSM) established in Ref.~\cite{Gallas:2009qp,Parganlija:2010fz,Janowski:2011gt,Eshraim:2012jv,Parganlija:2012fy,Janowski:2014ppa,Divotgey:2016pst}, in which vector and axial-vector mesons are incorporated in addition to the scalar and pseudo-scalar mesons,\footnote{Further extensions of the Linear Sigma Model are possible by including tetraquark states~\cite{Giacosa:2006tf,Fariborz:2008bd,Lakaschus:2018rki}, for example.} while global chiral symmetry and scale invariance is respected. The eLSM not only reproduces meson properties in vacuum such as masses and decay widths, but has also been proven to be useful for the investigation of nuclear matter as well \cite{Gallas:2011qp}.

In this study, the nucleons are introduced by the two-flavor Parity Doublet Model (PDM)~\cite{Detar:1988kn,Nemoto:1998um,Jido:1998av,Jido:1999hd,Jido:2001nt}, and nuclear matter is constructed in the one-loop approximation of the nucleon~\cite{Zschiesche:2006zj,Gallas:2011qp,Motohiro:2015taa,Suenaga:2017wbb} while the mesons are obtained from the eLSM. In our approach, not only the vacuum properties of nucleons but also the nuclear matter properties such as the saturation density, the binding energy per nucleon and the incompressibility are successfully reproduced. The PDM contains two types of nucleons, the nucleon and its chiral partner, and in the so-called {\it mirror assignment}, it is possible to construct a nucleon mass term without violating chiral symmetry. In other words, the PDM predicts the existence of a nucleon mass that does not originate from chiral symmetry breaking, which is the so-called {\it chiral invariant mass} ($M_0$). The existence of $M_0$ is also suggested by lattice calculations in the context of a parity doubling, however, its precise value is still under discussion~\cite{DeTar:1987ar,DeTar:1987xb,Aarts:2017rrl}. In this paper, we find the value of $M_0 \approx 0.8$ GeV to be preferable. 

This paper is organized as follows. In Sec.~\ref{sec:ELSM}, the three-flavor eLSM is introduced and the determined model parameters are shown provided by Ref.~\cite{Parganlija:2012fy}. In Sec.~\ref{sec:NuclearMatter}, we present the two-flavor PDM and construct nuclear matter by combining the eLSM and the PDM. In Sec.~\ref{sec:Results}, the remaining parameters are determined and numerical results of the density dependence of the meson masses are presented. Sec.~\ref{sec:EtaImprove} and Sec.~\ref{sec:Discussion} are devoted to the discussions and conclusions, respectively.

%%%%%%%%%%%%%%%%%%%%%

\section{Extended Linear Sigma Model (eLSM)}
\label{sec:ELSM}

In this paper, we present a comprehensive study of light mesons in nuclear matter by employing the three-flavor eLSM. This model was established in Ref.~\cite{Parganlija:2012fy}, and successfully reproduced meson properties in vacuum such as masses and decay widths. In this section we introduce the eLSM and present the determined parameters of the above reference.

The Lagrangian of the eLSM maintains a scale invariance except for current quark mass effects, in which a dilaton is responsible for the violation of scale invariance of QCD, which is given by
\begin{widetext}
\begin{eqnarray}
{\cal L}_{\rm eLSM} &=& {\cal L}_{\rm dil} + {\rm Tr}[(D_\mu\Phi)^\dagger(D^\mu\Phi)]-m_0^2\left(\frac{G}{G_0}\right)^2{\rm Tr}[\Phi^\dagger\Phi]-\lambda_1\left({\rm Tr}[\Phi^\dagger\Phi]\right)^2-\lambda_2{\rm Tr}[(\Phi^\dagger\Phi)^2] +{\rm Tr}[H(\Phi^\dagger+\Phi)]\nonumber\\
&& -\frac{1}{4}{\rm Tr}[L_{\mu\nu}L^{\mu\nu} + R_{\mu\nu}R^{\mu\nu}] + {\rm Tr}\left[\left(\frac{m_1^2}{2}\left(\frac{G}{G_0}\right)^2+\Delta\right)(L_\mu^2+R_\mu^2)\right] + c_1\left({\rm det}\Phi-{\rm det}\Phi^\dagger\right)^2 \nonumber\\
&& + i\frac{g_2}{2}\Big({\rm Tr}[L_{\mu\nu}[L^\mu, L^\nu] ] + {\rm Tr}[R_{\mu\nu}[R^\mu,R^\nu]]  \Big) \nonumber\\
&& + \frac{h_1}{2}{\rm Tr}[\Phi^\dagger\Phi]{\rm Tr}[L_\mu^2+R_\mu^2] + h_2{\rm Tr}[L_\mu^2\Phi\Phi^\dagger+R_\mu^2\Phi^\dagger\Phi] + 2h_3{\rm Tr}[L_\mu\Phi R^\mu \Phi^\dagger]  \nonumber\\
&& +g_3\Big({\rm Tr}[L_\mu L_\nu L^\mu L^\nu]+{\rm Tr}[R_\mu R_\nu R^\mu R^\nu] \Big)+g_4\Big({\rm Tr}[L_\mu L^\mu L_\nu L^\nu]+{\rm Tr}[R_\mu R^\mu R_\nu R^\nu] \Big) \nonumber\\
&& + g_5 {\rm Tr}[L_\mu L^\mu]{\rm Tr}[R_\nu R^\nu] + g_6\Big({\rm Tr}[L_\mu L^\mu]{\rm Tr}[L_\nu L^\nu] + {\rm Tr}[R_\mu R^\mu]{\rm Tr}[R_\nu R^\nu] \Big)\ , \label{eLSMOriginal}
\end{eqnarray}
where the meson nonets $\Phi$, $L_\mu$, and $R_\mu$ are
\begin{eqnarray}
\Phi &=& \frac{1}{\sqrt{2}}\left(
\begin{array}{ccc}
\frac{\sigma_N+a_0^0+i(\eta_N+\pi^0)}{\sqrt{2}} &a_0^++ i\pi^+& K_0^{*+}+iK^+ \\
a_0^-+i\pi^- & \frac{\sigma_N-a_0^0+i(\eta_N-\pi^0)}{\sqrt{2}}  & K_0^{*0} + iK^0 \\
K_0^{*-}+iK^-& \bar{K}_0^{*0} + i\bar{K}^0 & \sigma_S+i\eta_S \\
\end{array} 
\right) \ ,\nonumber\\ 
%%%%%%
L_\mu &=& \frac{1}{\sqrt{2}}\left(
\begin{array}{ccc}
\frac{\omega_N+\rho^0}{\sqrt{2}} +  \frac{f_{1N}+a_1^0}{\sqrt{2}}  & \rho^++a_1^+& K^{*+} + K_1^+ \\
\rho^- + a_1^- &\frac{\omega_N-\rho^0}{\sqrt{2}} +  \frac{f_{1N}-a_1^0}{\sqrt{2}}  & K^{*0} + K_1^0 \\
K^{*-} + K_1^- & \bar{K}^{*0} +\bar{K}_1^0 & \omega_S+f_{1S} \\
\end{array} 
\right)_\mu \ , \nonumber\\ 
%%%%%%
R_\mu &=& \frac{1}{\sqrt{2}}\left(
\begin{array}{ccc}
\frac{\omega_N+\rho^0}{\sqrt{2}} -  \frac{f_{1N}+a_1^0}{\sqrt{2}}  & \rho^+-a_1^+& K^{*+}-K_1^+ \\
\rho^--a_1^- &\frac{\omega_N-\rho^0}{\sqrt{2}} -  \frac{f_{1N}-a_1^0}{\sqrt{2}}  & K^{*0} - K_1^0 \\
K^{*-}-K_1^-& \bar{K}^{*0} -\bar{K}_1^0 & \omega_S-f_{1S} \\
\end{array} 
\right)_\mu \ .
\end{eqnarray}
\end{widetext}
$G$ is the dilaton field and its kinetic term and self-interaction terms are included in ${\cal L}_{\rm dil}$.\footnote{Explicitly, ${\cal L}_{\rm dil}$ is of the form~\cite{Rosenzweig:1981cu,Salomone:1980sp,Rosenzweig:1982cb,Migdal:1982jp,Gomm:1984zq,Gomm:1985ut}
\begin{eqnarray}
{\cal L}_{\rm dil} = \frac{1}{2}\partial_\mu G\partial^\mu G-\frac{1}{4}\frac{m_G^2}{G_0^2}\left(G^4{\rm ln}\frac{G}{G_0}-\frac{G^4}{4}\right)\ ,
\end{eqnarray}
with $G_0$ the dilaton mean field, and $m_G$ the dilaton mass which is matched by the trace anomaly of QCD.} The chiral transformation laws for $\Phi$, $L_\mu$, $R_\mu$ are
\begin{eqnarray}
\Phi \to g_L \Phi g_R^\dagger\ ,\ \ L_\mu \to g_L L_\mu g_L^\dagger\ ,\ \  R_{\mu}\to g_R R_\mu g_R^\dagger\ , \label{ChiralTrans}
\end{eqnarray}
where $g_L$ ($g_R$) is an element of $U(3)_L$ $(U(3)_R)$ chiral group. $L_{\mu\nu}$ and $R_{\mu\nu}$ in Eq.~(\ref{eLSMOriginal}) are the field strengths
\begin{eqnarray}
L_{\mu\nu} = \partial_\mu L_\nu-\partial_\nu L_\mu \ ,\ \  R_{\mu\nu} = \partial_\mu R_\nu-\partial_\nu R_\mu \ ,
\end{eqnarray}
representing the kinetic terms for the (axial-)vector mesons. The covariant derivative is $D_\mu \Phi = \partial_\mu\Phi -ig_1(L_\mu \Phi- \Phi R_\mu)$.\footnote{In fact, $L_\mu$ and $R_\mu$ are not ``gauge fields'' as understood by the transformation laws in Eq.~(\ref{ChiralTrans}), such that these vector mesons do not need to couple with $\Phi$ by a covariant derivative unlike in the context of Hidden Local Symmetry (HLS)~\cite{Harada:2003jx}. The remnant contributions are provided by $h_2$ and $h_3$ terms.} The matrices $H$ and $\Delta$ are responsible for the explicit breaking of chiral symmetry which take forms of $H= {\rm diag}(\frac{h_{0N}}{2},\frac{h_{0N}}{2}, \frac{h_{0S}}{\sqrt{2}})$ and $\Delta = {\rm diag}(\delta_N,\delta_{N}, \delta_{S})$, respectively.

In the following analysis, we will regard the dilaton as the $f_0(1710)$ whose mass is larger than the other light meson masses. Thus, the dilaton dynamics will be ignored in what follows, {\it i.e.}, the dilaton field in Eq.~(\ref{eLSMOriginal}) is simply replaced by its mean field: $G\to G_0$.  Also, we assume the large-$N_c$ suppression works well for interactions containing the spin-$1$ mesons, which allows us to drop the single-trace terms~\cite{tHooft:1973alw,Witten:1979kh}. Hence, the reduced three-flavor eLSM reads
\begin{eqnarray}
{\cal L}^{\rm red}_{\rm eLSM} &=&  {\rm Tr}[(D_\mu\Phi)^\dagger(D^\mu\Phi)]-m_0^2{\rm Tr}[\Phi^\dagger\Phi]-\lambda_1\left({\rm Tr}[\Phi^\dagger\Phi]\right)^2 \nonumber\\
&&-\lambda_2{\rm Tr}[(\Phi^\dagger\Phi)^2] +{\rm Tr}[H(\Phi^\dagger+\Phi)] -\frac{1}{4}{\rm Tr}[L_{\mu\nu}L^{\mu\nu} \nonumber\\
&&+ R_{\mu\nu}R^{\mu\nu}] + {\rm Tr}\left[\left(\frac{m_1^2}{2}+\Delta\right)(L_\mu^2+R_\mu^2)\right]  \nonumber\\
&&+ c_1\left({\rm det}\Phi-{\rm det}\Phi^\dagger\right)^2 + h_2{\rm Tr}[L_\mu^2\Phi\Phi^\dagger+R_\mu^2\Phi^\dagger\Phi]  \nonumber\\
&&+ 2h_3{\rm Tr}[L_\mu\Phi R^\mu \Phi^\dagger]  \nonumber\\
&& +g_{4p}\Big({\rm Tr}[L_\mu L_\nu L^\mu L^\nu]+{\rm Tr}[R_\mu R_\nu R^\mu R^\nu] \Big) \nonumber\\
&&+g_{4p}\Big({\rm Tr}[L_\mu L^\mu L_\nu L^\nu]+{\rm Tr}[R_\mu R^\mu R_\nu R^\nu] \Big) \ . \label{eLSM}
\end{eqnarray}
We note that the $g_2$ term is dropped as well, although this term can provide mass modifications to the spin-$1$ kaon sector in nuclear matter due to the $\omega_N$ mean field. We expect the correction is small since the $\omega_N$ mean field is suppressed in comparison to the $\sigma_N$ or $\sigma_S$ meson fields. We also note that due to the lack of information on the values for the four-point couplings of spin-$1$ mesons, we have taken $g_3=g_4 \equiv g_{4p}$ for simplicity.

The values of the model parameters determined in Ref.~\cite{Parganlija:2012fy} are listed in Table~\ref{tab:PeLSM}. These parameters are fixed in order to reproduce the masses and decay widths of light mesons in vacuum, with the mean fields of $\sigma_N$ and $\sigma_S$, $\hat{\phi}_N\equiv \langle\sigma_N\rangle_{\rm vac}$ and $\hat{\phi}_S\equiv\langle\sigma_S\rangle_{\rm vac}$,\footnote{Throughout this paper, we use a symbol ``$\hat{X}$'' for referring to a vacuum value of the quantity $X$} satisfying gap equations. The detailed procedure to fix the parameters are given in Ref.~\cite{Parganlija:2012fy}. According to this reference, the value of $\lambda_1$ cannot be fixed due to a large uncertainty of the $f_0$ (scalar-isoscalar) meson sector. Besides, as we have mentioned in the previous paragraph, the value of $g_{4p}$ remains to be determined. The parameters $\lambda_1$ and $g_{4p}$ will be determined by fitting them to nuclear matter properties in Sec.~\ref{sec:Parameters}. In fact, $g_{4p}$ can play a significant role in reproducing the incompressibility of nuclear matter~\cite{Zschiesche:2006zj}.

%%%%%%%%%%%%%%%%%%%%%%%%%
\begin{table}[thbp]
  \begin{tabular}{cc} \hline\hline 
Parameters in eLSM      &       Values   \\\hline
$C_1$ [GeV$^{2}$] & -0.9183\\
$C_2$ [GeV$^{2}$] & 0.4135\\
$c_1$ [GeV$^{-2}$] & 450.5\\
$\delta_N$ [GeV$^{2}$] & 0\\
$\delta_S$ [GeV$^{2}$] & 0.1511\\
$g_1$ & 5.8433 \\
$\hat{\phi}_{N}$ [GeV] & 0.1646 \\
$\hat{\phi}_{S}$ [GeV] & 0.1262 \\
$h_{0N}$ [GeV$^3$] & 0.001135\\
$h_{0S}$ [GeV$^3$] &0.02138 \\
$h_2$ & 9.880 \\
$h_3$ & 4.867  \\
$\lambda_2$ & 68.30
\\ \hline \hline
  \end{tabular}
\caption{Parameters extracted from the eLSM in Ref.~\cite{Parganlija:2012fy}. Here, $C_1=m_0^2+\lambda_1(\hat{\phi}_{N}^2+\hat{\phi}_{S}^2)$ and $C_2 = m_1^2$ with $\hat{\phi}_{N}$ and $\hat{\phi}_{S}$ being the VEV of $\sigma_N$ and $\sigma_S$ in the vacuum }
  \label{tab:PeLSM}
\end{table}
%%%%%%%%%%%%%%%%%%%%%%%%%

%%%%%%%%%%%%%%%%%%%%%

\section{Construction of nuclear matter}
\label{sec:NuclearMatter}

\subsection{Parity Doublet Model (PDM)}
\label{sec:PDM}

In this study, the meson masses in nuclear matter are determined by calculating the one-loop nucleon corrections to the meson mean fields. To this end, we combine the three-flavor eLSM and the two-flavor PDM. Although light mesons containing (anti-)strange quarks can in principle couple with nucleons, we do not include such interactions for the sake of a clear and transparent study. Hence, the Lagrangian of the PDM is given by~\cite{Detar:1988kn,Jido:2001nt}
\begin{widetext}
\begin{eqnarray}
{\cal L}_N &=& \bar{\psi}_{1r}(i\Slash{\partial}+\mu_B\gamma_0+g_V\Slash{\tilde{R}})\psi_{1r} + \bar{\psi}_{1l}(i\Slash{\partial}+\mu_B\gamma_0+g_V\Slash{\tilde{L}})\psi_{1l}+\bar{\psi}_{2r}(i\Slash{\partial}+\mu_B\gamma_0 + h_V \Slash{\tilde{L}})\psi_{2r} + \bar{\psi}_{2l}(i\Slash{\partial}+\mu_B\gamma_0+h_V\Slash{\tilde{R}})\psi_{2l} \nonumber\\
&& + \tilde{g}_{1V}\left({\rm Tr}[\tilde{R}_\mu]\bar{\psi}_{1r}\gamma^\mu\psi_{1r}+{\rm Tr}[\tilde{L}_\mu]\bar{\psi}_{1l}\gamma^\mu\psi_{1l} \right)+  \tilde{g}_{2V}\left({\rm Tr}[\tilde{R}_\mu]\bar{\psi}_{1l}\gamma^\mu\psi_{1l}+{\rm Tr}[\tilde{L}_\mu]\bar{\psi}_{1r}\gamma^\mu\psi_{1r} \right) \nonumber\\
&& + \tilde{h}_{1V}\left({\rm Tr}[\tilde{R}_\mu]\bar{\psi}_{2r}\gamma^\mu\psi_{2r}+{\rm Tr}[\tilde{L}_\mu]\bar{\psi}_{2l}\gamma^\mu\psi_{2l} \right)+  \tilde{h}_{2V}\left({\rm Tr}[\tilde{R}_\mu]\bar{\psi}_{2l}\gamma^\mu\psi_{2l}+{\rm Tr}[\tilde{L}_\mu]\bar{\psi}_{2r}\gamma^\mu\psi_{2r} \right) \nonumber\\
&& - M_0\left[\bar{\psi}_{1l}\psi_{2r}-\bar{\psi}_{1r}\psi_{2l}-\bar{\psi}_{2l}\psi_{1r}+\bar{\psi}_{2r}\psi_{1l}\right]  -k_1({\rm det} \tilde{\Phi}+{\rm det} \tilde{\Phi}^\dagger)\left[\bar{\psi}_{1l}\psi_{2r}-\bar{\psi}_{1r}\psi_{2l}-\bar{\psi}_{2l}\psi_{1r}+\bar{\psi}_{2r}\psi_{1l}\right]   \nonumber\\
&& -k_2\left({\rm det} \tilde{\Phi}-{\rm det} \tilde{\Phi}^\dagger\right)\left[\bar{\psi}_{1l}\psi_{2r}+\bar{\psi}_{1r}\psi_{2l}+\bar{\psi}_{2l}\psi_{1r}+\bar{\psi}_{2r}\psi_{1l}\right] \nonumber\\
&& -G_1\left[\bar{\psi}_{1r}\tilde{\Phi}^{\dagger}\psi_{1l}+\bar{\psi}_{1l}\tilde{\Phi}\psi_{1r}\right]-G_2\left[\bar{\psi}_{2r}\tilde{\Phi}^{\dagger}\psi_{2l}+\bar{\psi}_{2l}\tilde{\Phi}\psi_{2r}\right] \ , \label{LNStart}
\end{eqnarray}
\end{widetext}
in which $\psi_{1r(l)}$ is the {\it naive-assigned} nucleon and $\psi_{2r(l)}$ is the {\it mirror-assigned} one, {\it i.e.}, these nucleons transform under the $U(2)_L\times U(2)_R$ chiral transformation as
\begin{eqnarray}
\psi_{1r} \to \tilde{g}_R\psi_1 \  &,&\ \psi_{1l} \to \tilde{g}_L\psi_{1l}\ , \nonumber\\
\psi_{2r} \to \tilde{g}_L \psi_2\ &,&\ \psi_{2l} \to \tilde{g}_R\psi_{2r}\ ,
\end{eqnarray}
with $\tilde{g}_{L} \in U(2)_{L}$ and $\tilde{g}_R\in U(2)_R$. $\tilde{\Phi}$, $\tilde{R}_\mu$, and $\tilde{L}_\mu$ are two-flavor projected light meson fields given by
\begin{eqnarray}
\tilde{\Phi} &=& \sigma_N + i\pi^a\tau^a\ , \nonumber\\
\tilde{V}_\mu &=& \frac{\tilde{L}_\mu+\tilde{R}_\mu}{2} = \frac{1}{2}(
{\omega}_N+{\rho}^a\tau^a)_\mu \ ,\nonumber\\
\tilde{A}_\mu &=& \frac{\tilde{L}_\mu-\tilde{R}_\mu}{2}= \frac{1}{2} (
{f}_{1N}+{a}_1^a\tau^a)_{\mu} \ ,
\end{eqnarray}
with $\tau^a$ the Pauli matrices. In Eq.~(\ref{LNStart}), $\mu_B$ is a baryon number chemical potential introduced to access finite baryon density. Unfamiliar terms are the $k_1$ and $k_2$ ones which include determinants of the multiplets in flavor space. Although the mass dimensions of $k_1$ and $k_2$ are [GeV$^{-1}$] in the two-flavor case, these terms are allowed by the $U(1)_A$ axial anomaly in principle. Especially the $k_2$ term is essential to reproduce the decay width of $N^*(1535)\to N\eta$ decay in vacuum~\cite{Olbrich:2017fsd}. The $k_1$ term provides an additional contribution to the nucleon masses as will be observed soon. Under chiral symmetry breaking at finite density, $\sigma_N$, $\sigma_S$, and the time-component of $\omega_N^\mu$ possess the mean field values:
\begin{eqnarray}
\phi_N \equiv \langle\sigma_N\rangle\ , \ \ \phi_S \equiv \langle\sigma_S\rangle\ , \ \ \bar{\omega}_N \equiv \langle\omega_N^{\mu=0}\rangle\ , \label{MeanField}
\end{eqnarray}
respectively ($\phi_S$ enters through the eLSM in Eq.~(\ref{eLSM})). In vacuum these values are reduced to $\phi_N\to\hat{\phi}_N$, $\phi_S \to \hat{\phi}_S$ and $\bar{\omega}_N\to0$. Note that the trace terms proportional to $\tilde{g}_{1V}$, $\tilde{h}_{1V}$, $\tilde{g}_{2V}$, and $\tilde{h}_{2V}$ allowed by the chiral symmetry are included to provide a difference between the $\rho NN$ and $\omega_{N}NN$ couplings, which make the density dependences of the $\rho$ and $\omega_N$ masses differ.

In Lagrangian~(\ref{LNStart}), although $\psi_{1r(1l)}$ and $\psi_{2r(2l)}$ are convenient to observe the chiral symmetric properties of the Lagrangian, these fields are not mass eigenstates. The mass eigenstates $N_+$ and $N_-$ are obtained by introducing a mixing angle $\theta$ as 
\begin{eqnarray}
\left(
\begin{array}{c}
N_+ \\
N_- \\
\end{array}
\right) = \left(
\begin{array}{cc}
{\rm cos}\, \theta & \gamma_5{\rm sin}\, \theta \\
-\gamma_5{\rm sin}\, \theta & {\rm cos}\, \theta \\
\end{array}
\right)\left(
\begin{array}{c}
\psi_1 \\
\psi_2\\
\end{array}
\right)\ ,
\end{eqnarray}
with $\theta$ satisfying
\begin{eqnarray}
{\rm tan}\, 2\theta &=& \frac{2\left(M_0+\frac{k_1}{2}\phi_N^2\right) }{(G_1+G_2)\phi_N}\ ,\nonumber\\ \nonumber\\
{\rm cos}\, 2\theta &=& \frac{(G_1+G_2)\phi_N}{\sqrt{(G_1+G_2)^2\phi_N^2+4\left(M_0+\frac{k_1}{2}\phi_N^2\right)^2}}\ ,\nonumber\\ \nonumber\\
{\rm sin}\, 2\theta &=& \frac{2\left(M_0+\frac{k_1}{2}\phi_N^2\right)}{\sqrt{(G_1+G_2)^2\phi_N^2+4\left(M_0+\frac{k_1}{2}\phi_N^2\right)^2}} \ ,  \label{MixAngle}
\end{eqnarray}
and the corresponding mass eigenvalues are
\begin{eqnarray}
m_\pm &=&  \frac{1}{2}\Big(\sqrt{(G_1+G_2)^2\phi_N^2+4\left(M_0+\frac{k_1}{2}\phi_N^2\right)^2} \nonumber\\
&& \mp(G_2-G_1)\phi_N\Big) \ , \label{MassFormula}
\end{eqnarray}
for $N_\pm$ (double-sign correspondence). $N_+$ is a positive-parity state while $N_-$ is a negative-parity state, then, we assign $N_+$ to the nucleon $N(939)$ and $N_-$ the $N^*(1535)$. Eqs.~(\ref{MixAngle}) and~(\ref{MassFormula}) show that the direct $U(1)_A$ anomaly correction to the nucleons ($k_1$ term) can modify both the mixing angle $\theta$ and the mass eigenvalues $m_\pm$. At the point of chiral restoration $\phi_N=0$, Eq.~(\ref{MassFormula}) yields $m_\pm \to M_0$ which shows that the nucleon masses can be generated without chiral symmetry breaking.\footnote{According to Ref. \cite{Casher:1979vw} deconfinement implies restoration of chiral symmetry and therefore no hadrons should exist in the chiral limit. However, this argument does not generalize to finite density as argued by Glozman \cite{Glozman:2009sa} and supported by lattice investigations \cite{Glozman:2012fj, Suganuma:2017syi}.} This is why $M_0$ is often refereed to as a {\it chiral invariant mass}.

By assuming $\tilde{g}_{1V} = \tilde{h}_{1V} = \tilde{g}_{2V} = \tilde{h}_{2V} \equiv \tilde{g}$ (but still $g_V$ and $h_V$ are not identical) for simplicity, the Lagrangian~(\ref{LNStart}) is rewritten into
\begin{widetext}
\begin{eqnarray}
{\cal L}_N &=&  \bar{N}_+i\Slash{\partial}N_++\bar{N}_-i\Slash{\partial}N_--m_+\bar{N}_+N_+-m_-\bar{N}_-N_-  \nonumber\\
&& + \big(g_V{\rm cos}^2\theta+h_V{\rm sin}^2\theta\big)\bar{N}_+\Slash{\tilde{V}}N_+ +\big(g_V{\rm cos}^2\theta-h_V{\rm sin}^2\theta\big)\bar{N}_+\Slash{\tilde{A}}\gamma_5N_+ \nonumber\\ 
&& -(g_V-h_V){\rm sin}\, \theta\, {\rm cos}\, \theta\bar{N}_+\Slash{\tilde{V}}\gamma_5 N_--(g_V+h_V){\rm sin}\, \theta\, {\rm cos}\, \theta\bar{N}_+\Slash{\tilde{A}} N_- \nonumber\\
&& -(g_V-h_V){\rm sin}\, \theta\, {\rm cos}\, \theta\bar{N}_-\Slash{\tilde{V}}\gamma_5 N_+-(g_V+h_V){\rm sin}\, \theta\, {\rm cos}\, \theta\bar{N}_-\Slash{\tilde{A}} N_+ \nonumber\\
&& + \big(g_V{\rm sin}^2\theta+h_V{\rm cos}^2\theta\big)\bar{N}_-\Slash{\tilde{V}}N_- +\big(g_V{\rm sin}^2\theta-h_V{\rm cos}^2\theta\big)\bar{N}_-\Slash{\tilde{A}}\gamma_5N_- \nonumber\\
%%%%%%
&&+2\tilde{g}\bar{N}_+\Slash{\omega}_N N_+ +2\tilde{g}\bar{N}_-\Slash{\omega}_N N_- \nonumber\\
%%%%%%%%
&& -k_1({\rm det} {\Phi}+{\rm det} {\Phi}^\dagger)_{\rm fl} \left\{{\rm sin}\, 2\theta\, \bar{N}_+N_++{\rm cos}\, 2\theta\, \bar{N}_+\gamma_5 N_--{\rm cos}\, 2\theta\, \bar{N}_-\gamma_5 N_++{\rm sin}\, 2\theta\, \bar{N}_-N_- \right\} \nonumber\\
&&-k_2({\rm det} {\Phi}-{\rm det}{\Phi}^\dagger)_{\rm fl}  \left\{{\rm sin}\, 2\theta\, \bar{N}_+\gamma_5N_++{\rm cos}\, 2\theta\, \bar{N}_+ N_--{\rm cos}\, 2\theta\, \bar{N}_- N_++{\rm sin}\, 2\theta\, \bar{N}_-\gamma_5N_- \right\} \nonumber\\
&&- g_{NN\sigma}\bar{N}_+(\sigma_N+a_0^a\tau^a) N_+-g_{NN\pi}\bar{N}_+i\gamma_5(\eta_N+\pi^a\tau^a) N_+ \nonumber\\
&& +g_{NN^*\sigma}\bar{N}_+\gamma_5(\sigma_N+a_0^a\tau^a) N_-+g_{NN^*\pi}\bar{N}_+i(\eta_N+\pi^a\tau^a) N_- \nonumber\\
&&-g_{NN^*\sigma}\bar{N}_-\gamma_5(\sigma_N+a_0^a\tau^a) N_+-g_{NN^*\pi}\bar{N}_-i(\eta_N+\pi^a\tau^a) N_+ \nonumber\\
&&-g_{N^*N^*\sigma}\bar{N}_-(\sigma_N+a_0^a\tau^a) N_--g_{N^*N^*\pi}\bar{N}_-i\gamma_5(\eta_N+\pi^a\tau^a) N_-  \ , \label{PDMThree}
\end{eqnarray}
\end{widetext}
in terms of $N_+$ and $N_-$, with
\begin{eqnarray}
({\rm det}\tilde{\Phi} + {\rm det}\tilde{\Phi}^\dagger)_{\rm fl} &=& \frac{1}{2}(2\phi_N\sigma_N+\sigma_N^2 \nonumber\\
&&-\eta_N^2-a_0^{a}a_0^a+\pi^a\pi^a) \ ,\nonumber\\
({\rm det}\tilde{\Phi}-{\rm det} \tilde{\Phi}^\dagger)_{\rm fl} &=& i\phi_N\eta_N\ ,
\end{eqnarray}
and
\begin{eqnarray}
g_{NN\sigma} &=& -\frac{G_2-G_1}{2}+\frac{G_1+G_2}{2}{\rm cos}\, 2\theta \ ,\nonumber\\
g_{NN\pi} &=& \frac{G_1+G_2}{2}+\frac{G_1-G_2}{2}{\rm cos}\, 2\theta \ ,\nonumber\\
g_{NN^*\sigma} &=& \frac{G_1+G_2}{2}{\rm sin}\, 2\theta\ , \nonumber\\
g_{NN^*\pi} &=&- \frac{G_2-G_1}{2}{\rm sin}\, 2\theta\ , \nonumber\\
g_{N^*N^*\sigma} &=& \frac{G_2-G_1}{2}+\frac{G_1+G_2}{2}{\rm cos}\, 2\theta \ ,\nonumber\\
g_{N^*N^*\pi} &=& -\frac{G_1+G_2}{2}-\frac{G_2-G_1}{2}{\rm cos}\, 2\theta\ .
\end{eqnarray}
%%%%%%%%%%%%%%%%%%%%%%%%%
\begin{table}[thbp]
  \begin{tabular}{cc} \hline\hline 
Inputs in the vacuum      &       Values   \\\hline
$\hat{m}_+$ [GeV] & 0.939\\
$\hat{m}_-$ [GeV] & 1.535\\
$\hat{g}_A^{N_+}$& 1.267 \\
$\hat{g}_A^{N_-}$& 0.2$\pm$0.3 ~\cite{Takahashi:2008fy} \\
$\hat{\Gamma}_{N^*(1535)\to N\eta}$ [GeV] & 0.065\\ \hline \hline
  \end{tabular}
\caption{Input parameters in terms of the vacuum properties of the nucleons.}
  \label{tab:InputVac}
\end{table}
%%%%%%%%%%%%%%%%%%%%%%%%%

The parameters in the PDM together with the remaining ones in the eLSM are fixed by both the vacuum properties of nucleons and nuclear matter properties. In terms of the vacuum properties, input parameters are the nucleon mass ($\hat{m}_+$), the $N^*(1535)$ mass ($\hat{m}_-$), the axial charges ($\hat{g}_A^{N_\pm}$), and the $N^*\to N\eta$ decay width ($\hat{\Gamma}_{N^*(1535)\to N\eta}$), which are summarized in Table~\ref{tab:InputVac} (Recall the symbol ``$\hat{X}$'' stands for a vacuum value of the quantity $X$). In our model, the axial charges and $N^*(1535)\to N \eta$ decay width are calculated as~\cite{Gallas:2009qp}
\begin{eqnarray}
\hat{g}_{A}^{N_+} &=& \frac{\hat{\phi}_N}{\hat{m}_+} \times\nonumber\\
&&  \left(\hat{g}_{NN\pi}+\big(g_V{\rm cos}^2\hat{\theta}-h_V{\rm sin}^2\theta\big)\frac{g_1}{\hat{m}_{a_1}^2}\hat{\phi}_N\hat{m}_+\right) \ ,\nonumber\\
\hat{g}_A^{N_-} &=& \frac{\hat{\phi}_N}{ \hat{m}_-}  \times\nonumber\\
&& \left(\hat{g}_{N^*N^*\pi}+\big(g_V{\rm sin}^2\hat{\theta}-h_V{\rm cos}^2\hat{\theta}\big)\frac{g_1}{\hat{m}_{a_1}^2}\hat{\phi}_N\hat{m}_-\right) \ , \nonumber\\
\end{eqnarray}
and
\begin{eqnarray}
\hat{\Gamma}_{N^*(1535)\to N\eta} = \frac{1}{8\pi}\frac{|\vec{p}_\eta|}{\hat{m}_-^2}\hat{G}^2\big[(\hat{m}_++\hat{m}_-)^2-\hat{m}_\eta^2\big] \ , \label{DecayWidth}
\end{eqnarray}
with
\begin{eqnarray}
|\vec{p}_\eta| &=& \frac{\sqrt{[\hat{m}_-^2-(\hat{m}_++\hat{m}_\eta)^2][\hat{m}_-^2-(\hat{m}_+-\hat{m}_\eta)^2]}}{2\hat{m}_-} \ , \nonumber\\
\hat{G} &=& -\hat{Z}_{\eta_N}\left(k_2\hat{\phi}_{N}{\rm cos}\, 2\hat{\theta}-\hat{g}_{NN^*\pi}\right){\rm cos}\, \hat{\theta}_\eta\ ,
\end{eqnarray}
respectively. Here, the renormalization factor $\hat{Z}_{\eta_N}$ and the mixing angle $\hat{\theta}_{\eta}$ are given in Eq.~(\ref{ZEtaVac}) and Eq.~(\ref{EtaMix}), respectively. As already mentioned, the parameter $k_2$ is crucial to fit the $N^*(1535)\to N\eta$ decay in Eq.~(\ref{DecayWidth})~\cite{Olbrich:2017fsd}. The list of the determined parameters after fitting nuclear matter properties will be summarized in Table~\ref{tab:ParametersPD} in Sec.~\ref{sec:Parameters}.

\subsection{Nuclear matter}
\label{sec:Matter}
In this subsection we construct nuclear matter by combining the eLSM and the PDM provided in Sec.~\ref{sec:ELSM} and Sec.~\ref{sec:PDM}, and fit the remaining parameters. Here, nuclear matter is constructed by the one-loop approximation of the nucleon combined with the meson mean fields in Eq.~(\ref{MeanField}). Therefore, the grand potential (per volume) reads~\footnote{Here, contributions from $N_-$ are absent. This is true as far as we stick to lower density $\rho_B\lesssim 2\rho_0$ ($\rho_0$ is the normal nuclear density).}
\begin{widetext}
\begin{eqnarray}
\Omega/V &=& -\frac{1}{4\pi^2}\Bigg\{\frac{2}{3}\sqrt{k_F^{2}+m_+^2}k_F^{3}-\sqrt{k_F^{2}+m_+^2}k_F m_+^2+m_+^4{\rm ln}\left(\frac{k_F+\sqrt{k_F^{2}+m_+^2}}{m_+}\right)\Bigg\} \nonumber\\
&&+\left(\frac{m_0^2}{2}(\phi_N^2+\phi_S^2) + \frac{\lambda_1}{4}(\phi_N^2+\phi_S^2)^2+\frac{\lambda_2}{8}(\phi_N^4+2\phi_S^4)-h_{0N}\phi_N - h_{0S}\phi_S  -\frac{m_{\omega_N}^2}{2}\bar{\omega}^2_N -\frac{g_{4p}}{2}\bar{\omega}_N^4 \right)\nonumber\\
&&-\left(\frac{m_0^2}{2}(\hat{\phi}_{N}^2+\hat{\phi}_{S}^2) + \frac{\lambda_1}{4}(\hat{\phi}_{N}^2+\hat{\phi}_{S}^2)^2 + \frac{\lambda_2}{8}(\hat{\phi}_{N}^4+2\hat{\phi}_{S}^4) - h_{0N}\hat{\phi}_{N} - h_{0S}\hat{\phi}_{S} \right) \ , \label{OmegaPDM}
\end{eqnarray}
\end{widetext}
in which the effective chemical potential is 
\begin{eqnarray}
\mu_B^* \equiv \mu_B-g_\omega \bar{\omega}_N\ ,
\end{eqnarray}
with
\begin{eqnarray}
g_\omega=-\frac{1}{2}(g_V{\rm cos}^2\theta+h_V{\rm sin}^2\theta)-2\tilde{g}\ . \label{OmegaNN}
\end{eqnarray}
The Fermi momentum $k_F$ is defined via the relation $\mu_B^* = \sqrt{k_F^2+m_+^2}$, and the baryon number density $\rho_B$ is given by $\rho_B = \frac{2}{3\pi^2}k_F^3$. Note that the vacuum contributions to the grand potential have been subtracted to measure the thermodynamic quantities properly. The grand potential in Eq.~(\ref{OmegaPDM}) and gap equations with respect to $\phi_N$, $\phi_S$ and $\bar{\omega}_N$,
\begin{eqnarray}
\frac{\partial\Omega}{\partial\phi_N} = 0\ , \ \ \frac{\partial\Omega}{\partial\phi_S} = 0\ , \ \ \frac{\partial\Omega}{\partial\omega_N} = 0 \, , \label{GapEq}
\end{eqnarray}
are essential to get nuclear matter quantities and determine the density dependence of the meson masses consistently with the vacuum ones. We should note, in the current approach, the $\omega NN$ coupling ($g_\omega$) in Eq.~(\ref{OmegaNN}) depends on the density via the mixing angle $\theta$ because we have assumed $g_V\neq h_V$.

In this study, the saturation condition at the normal nuclear density $\rho_0=0.16$ fm$^{-3}$: $\frac{\partial}{\partial\rho_B}\left(\frac{E}{N}\right)|_{\rho_0}=0$ ($E$ is the total energy and $N$ is the mass number), the binding energy per nucleon $\frac{E}{N}|_{\rho_0}-\hat{m}_+=-16$ MeV, and the incompressibility $K=0.24$ GeV are chosen as input parameters by nuclear matter properties. First, the saturation condition reads
\begin{eqnarray}
\frac{\partial}{\partial \rho_B}\left(\frac{E}{N}\right) \Big|_{\rho_0} =\frac{P}{\rho_B^2}  \Big|_{\rho_0}= 0\ ,
\end{eqnarray}
in which $\rho_B=N/V$ have been utilized, with the help of simple thermodynamic relations
\begin{eqnarray}
E = -PV+\mu_B N\ , \label{TRel1}
\end{eqnarray}
and
\begin{eqnarray}
dE = \mu_B dN\ , \label{TRel2}
\end{eqnarray}
with $dV=0$. Thus, we arrive at
\begin{eqnarray}
P|_{\rho_0} = 0\ . \label{SatuP}
\end{eqnarray}
The pressure of the medium is simply defined by $P \equiv -\Omega/V$ with Eq.~(\ref{OmegaPDM}). Next, using Eq.~(\ref{SatuP}), the condition for the binding energy per nucleon is reduced to
\begin{eqnarray}
\frac{E}{N}\Big|_{\rho_0} = \mu_B |_{\rho_0}= 0.923\ {\rm GeV}\ , \label{BindingE}
\end{eqnarray}
together with Eq.~(\ref{TRel1}) and $\hat{m}_+= 0.939$ GeV.
Finally, again by using Eqs.~(\ref{TRel1}),~(\ref{TRel2}) and~(\ref{SatuP}), we obtain the incompressibility:
\begin{eqnarray}
K =9\rho_0^2\frac{\partial^2}{\partial\rho_B^2}\left(\frac{E}{N}\right)\Big|_{\rho_0} =
9\rho_0\left(\frac{\partial\mu_B}{\partial\rho_B}\right)_V\Big|_{\rho_0} = 0.24\ {\rm MeV}\ .\nonumber\\ \label{Inc}
\end{eqnarray}
The input parameters in terms of the nuclear matter properties are summarized in Table~\ref{tab:InputMedium}.
%%%%%%%%%%%%%%%%%%%%%%%%%
\begin{table}[thbp]
  \begin{tabular}{cc} \hline\hline 
Inputs in nuclear matter      &       Values   \\\hline
$P|_{\rho_0}$ & 0\\
$\mu_B|_{\rho_0}$ [GeV] & 0.923\\
$K$ [GeV]& 0.24 \\ \hline \hline
  \end{tabular}
\caption{Input parameters in terms of the nuclear matter properties.}
  \label{tab:InputMedium}
\end{table}
%%%%%%%%%%%%%%%%%%%%%%%%%

%%%%%%%%%%%%%%%%%%%%%%%%%
\begin{table*}[thbp]
  \begin{tabular}{cc||cccccccc} \hline\hline 
$M_0$ [GeV] &  $\tilde{k}_1$ & $G_1$ & $G_2$  & $\tilde{k}_2$ & $g_V$ & $h_V$ & $\tilde{g}$ &  $\lambda_1$ & $g_{4p}$ \\\hline
0.8 & 5 & -0.2924 & 3.329 & -19.60 & 21.10 & 8.063 & -6.333 & -22.73 & 39.09 \\ \hline
0.8 & 0 & 3.922 & 7.542  & -0.2467  & 3.847 & -9.186 & -2.623& -22.98 & 42.01 \\ \hline
0.8 & -5 & 5.324 & 8.945 & 0.8113 & 2.625 & -10.41 & -2.668 & -22.67 & 2.442 \\ \hline 
0.7 & 5 & 1.488 & 5.109 & -4.392 & 8.432 & -4.601 & -4.350 & -22.54 & 114.4 \\ \hline
0.7 & 0 & 4.386& 8.007 & 0.1265 & 3.381 & -9.652 & -3.683   & -22.84 & 171.9 \\ \hline
 0.7 & -5 & 5.497 & 9.118 & 0.9509 & 2.506 & -10.53 & -3.537 & -22.48 & 106.7 \\ \hline \hline
  \end{tabular}
\caption{
Determined remaining parameters with given $M_0$ and $k_1$. In this table, we define dimensionless quantities $\tilde{k}_1 = k_1\hat{\phi}_{N}$ and $\tilde{k}_2=k_2\hat{\phi}_{N}$. }
  \label{tab:ParametersPD}
\end{table*}
%%%%%%%%%%%%%%%%%%%%%%%%%

%%%%%%%%%
\begin{figure}[thbp]
\centering
\includegraphics*[scale=0.42]{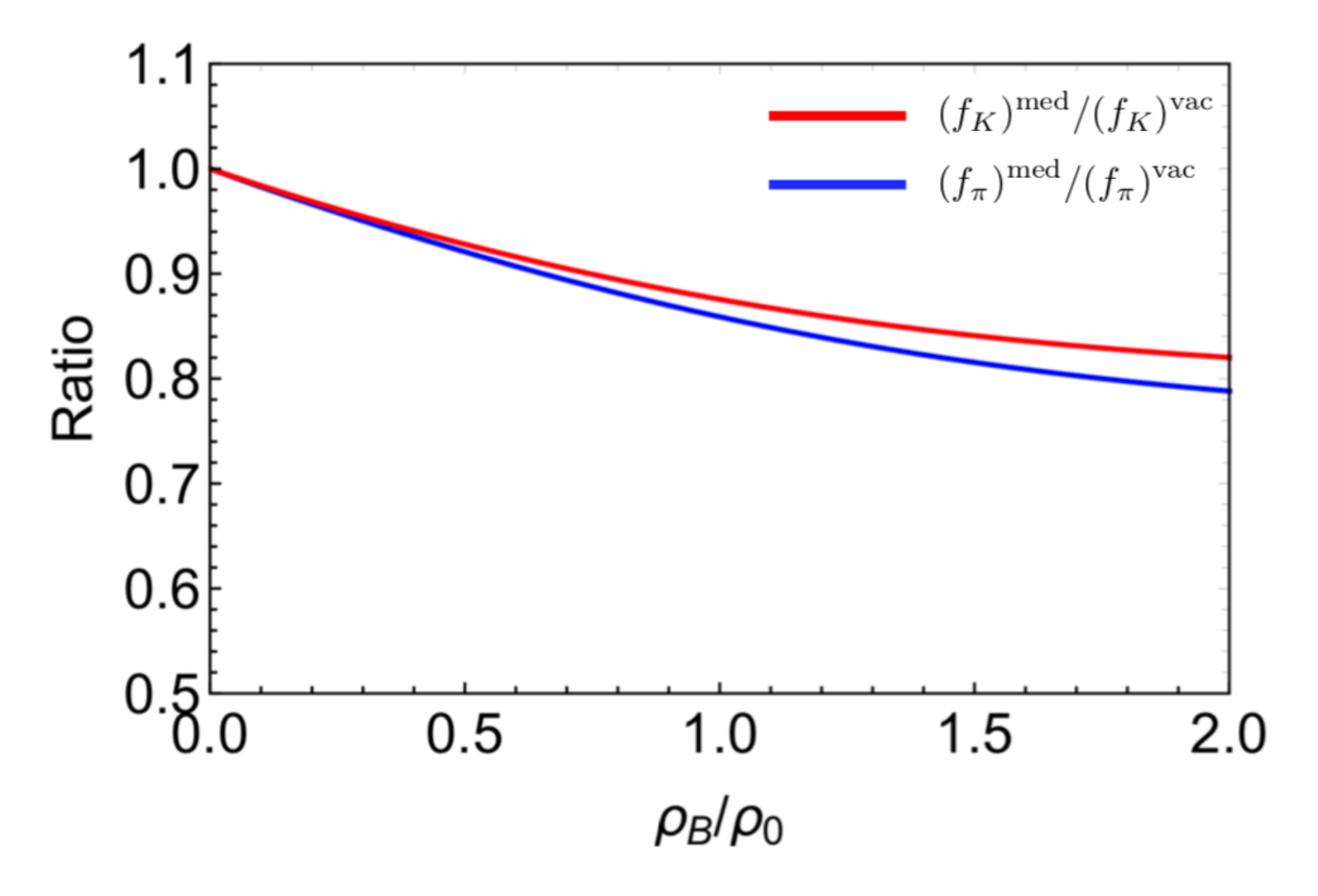}
\caption{(color online) The density dependence of $(f_\pi)^{\rm med}/(f_\pi)^{\rm vac}$ and $(f_K)^{\rm med}/(f_K)^{\rm vac}$ for $M_0=0.8$ GeV and $k_1=0$.}
\label{fig:FPiFK}
\end{figure}
%%%%%%%%%

%%%%%%%%%
\begin{figure*}[thbp]
\centering
\includegraphics*[scale=0.4]{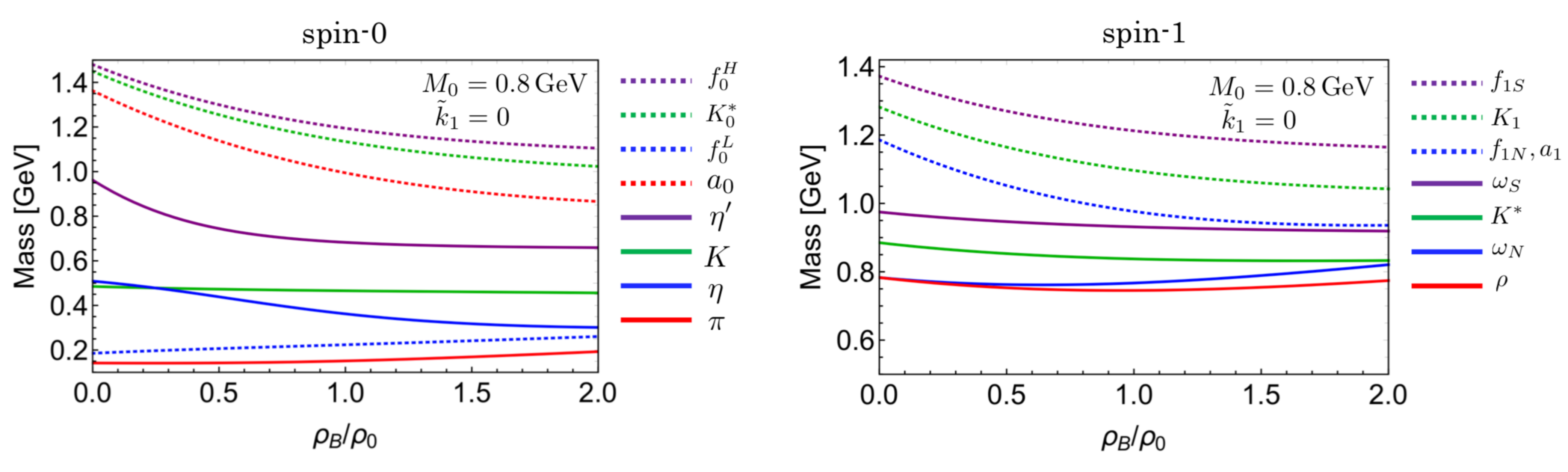}
\caption{(color online) The density dependence of spin-$0$ (left) and spin-$1$ (right) meson masses with $M_0=0.8$ GeV and $\tilde{k}_1=0$.}
\label{fig:M0800K10}
\end{figure*}
%%%%%%%%%
%%%%%%%%%
\begin{figure*}[thbp]
\centering
\includegraphics*[scale=0.4]{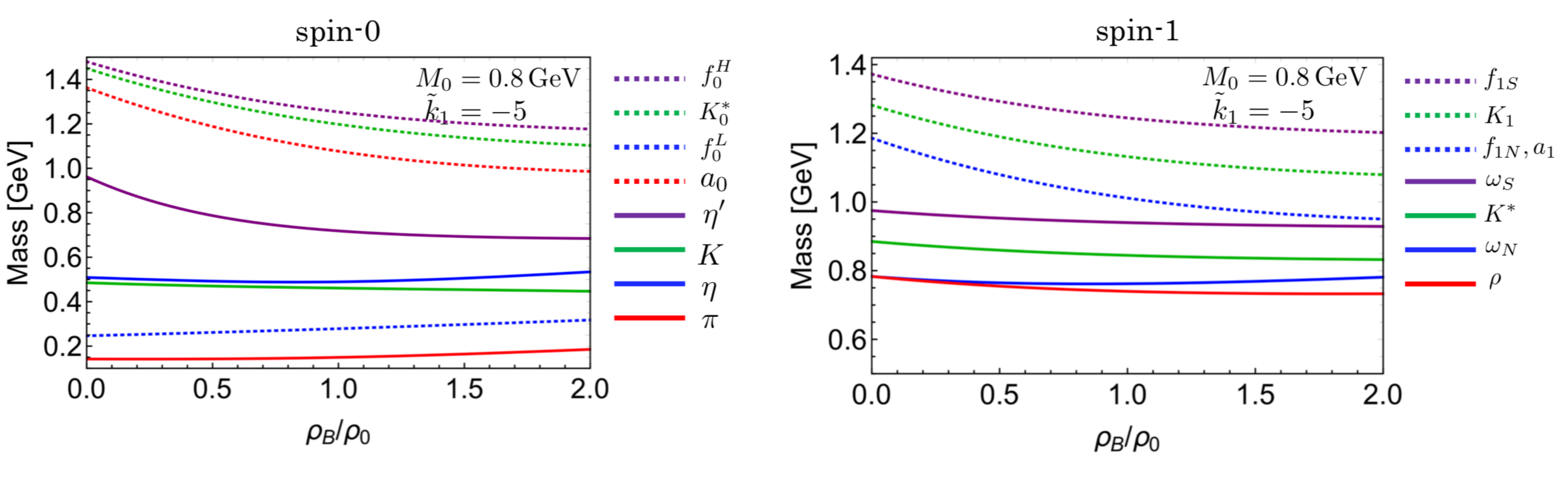}
\caption{(color online) The density dependence of spin-$0$ (left) and spin-$1$ (right) meson masses with $M_0=0.8$ GeV and $\tilde{k}_1=-5$.}
\label{fig:M0800K1-3}
\end{figure*}
%%%%%%%%%

%%%%%%%%%
\begin{figure*}[thbp]
\centering
\includegraphics*[scale=0.4]{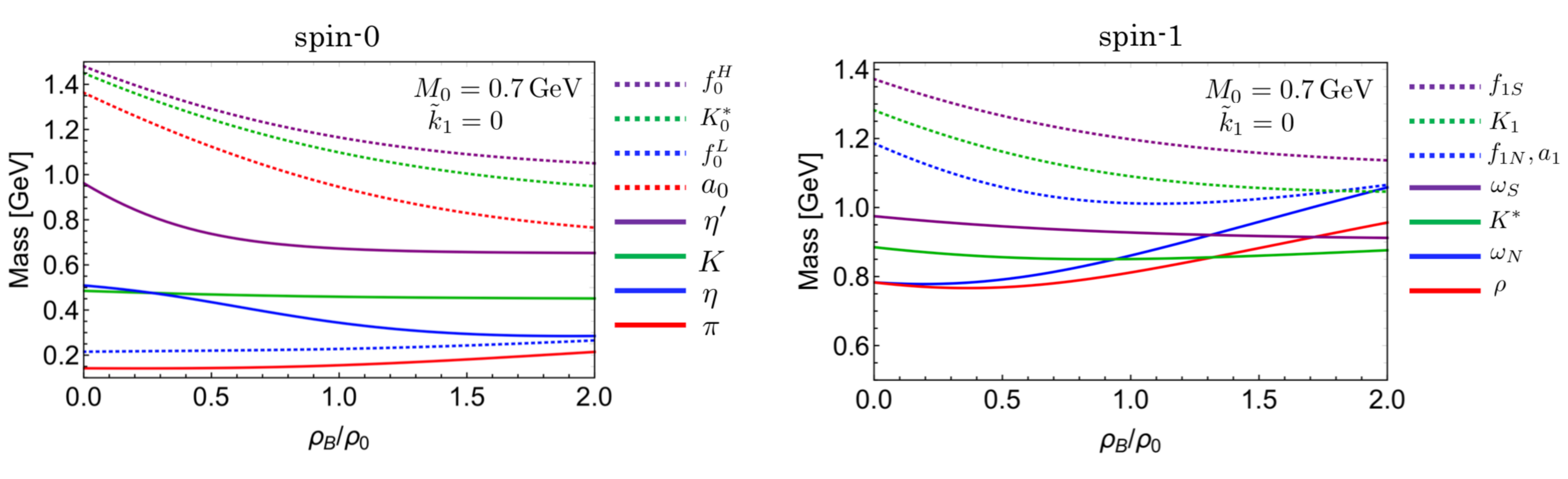}
\caption{(color online) The density dependence of spin-$0$ (left) and spin-$1$ (right) meson masses with $M_0=0.7$ GeV and $\tilde{k}_1=0$.}
\label{fig:M0700K10}
\end{figure*}
%%%%%%%%%
%%%%%%%%%
\begin{figure*}[thbp]
\centering
\includegraphics*[scale=0.4]{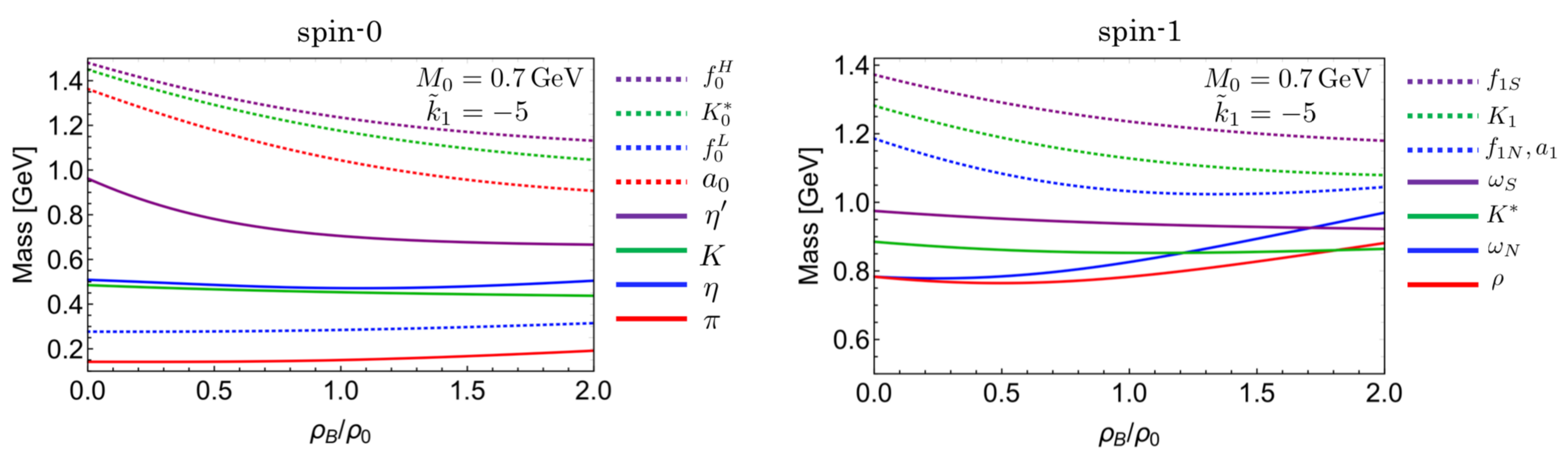}
\caption{(color online) The density dependence of spin-$0$ (left) and spin-$1$ (right) meson masses with $M_0=0.7$ GeV and $\tilde{k}_1=-5$.}
\label{fig:M0700K1-5}
\end{figure*}
%%%%%%%%%

\section{Results}
\label{sec:Results}

\subsection{Parameter determination}
\label{sec:Parameters}
Before showing numerical results of meson mass modifications in nuclear matter, we determine the remaining model parameters by fitting the inputs in Table~\ref{tab:InputVac} and Table~\ref{tab:InputMedium}. Our procedure leaves two free parameters. Hence, we select $M_0$ (chiral invariant mass) and $k_1$ (the strength of direct $U(1)_A$ anomaly effect to the nucleons) as free parameters.

The fixed parameters with a given set of $M_0$ and $k_1$ are summarized in Table~\ref{tab:ParametersPD}. Due to the strong constraints by nuclear matter properties, the allowed range of $M_0$ is $0.6\, {\rm GeV} \lesssim M_0\lesssim 0.8\, {\rm GeV}$. In terms of $k_2$, we could find two solutions since the formula to calculate the $N^*(1535) \to N\eta$ decay width in Eq.~(\ref{DecayWidth}) includes a quadratic term of $k_2$. Here, we pick up the solution of which the absolute value is smaller as done in Ref.~\cite{Olbrich:2017fsd}. Although the value of $\hat{g}_A^{N_-}$ includes a large uncertainty as given in Table~\ref{tab:InputVac}, we fix $\hat{g}_A^{N_-}=0.2$ here since the results are largely insensitive to a change of $\hat{g}_A^{N_-}$ within its error range. We should note that when the four-point interaction among spin-$1$ mesons ($g_{4p}$ term in Eq.~(\ref{eLSM})) is absent, we fail to reproduce the incompressibility. We assume the sign of $g_{4p}$ should be positive. Otherwise, we could find a nonzero $\bar{\omega}_N$ even in the vacuum which is forbidden by the Lorentz invariance. Furthermore, to determine the value of $\tilde{g}$, we have chosen a solution in such a way that the value of $\bar{\omega}_N$ is always positive. We emphasize that we could confirm the first order liquid-gas phase transition takes place at $\mu_B=0.923$ GeV~\cite{Pochodzalla:1995xy}.

A simple way to define the pion and kaon decay constants is~\cite{Son:2001ff}
\begin{eqnarray}
(f_\pi)^{\rm med} = \frac{\phi_N}{{Z}_\pi}\ , \ \ (f_K)^{\rm med} = \frac{\sqrt{2}\phi_S + \phi_N}{2{Z}_K}\ , \label{FPiFKDensDep}
\end{eqnarray}
as a naive extension of the vacuum ones: $(f_\pi)^{\rm vac}=\hat{\phi}_N/
\hat{Z}_\pi$ and $(f_K)^{\rm vac} = (\sqrt{2}\hat{\phi}_S+\hat{\phi}_N)/(2\hat{Z}_K)$ [$Z_\pi$, $Z_K$, $\hat{Z}_\pi$, and $\hat{Z}_K$ are renormalization factors defined in Eqs.~(\ref{ZPiZK}) and~(\ref{ZFactorsVac})]. By employing Eq.~(\ref{FPiFKDensDep}), the density dependences of $(f_\pi)^{\rm med} $ and $(f_K)^{\rm med}$ for $M_0=0.8$ GeV and $k_1=0$ are depicted in Fig.~\ref{fig:FPiFK}. This figure clearly shows the partial restoration of chiral symmetry at finite baryon density~\cite{Cohen:1991nk,Birse:1994cz}. The reduction ratios of $f_\pi$ and $f_K$ to the vacuum ones are $(f_\pi)^{\rm med}/(f_\pi)^{\rm vac}\approx 85 \%$ and $(f_K)^{\rm med}/(f_K)^{\rm vac}\approx 87 \%$, respectively.

\subsection{Numerical results}
\label{sec:NResults}
Here, we calculate self-energies of scalar, pseudo-scalar, vector, and axial-vector mesons, and show the resultant mass modifications in nuclear matter. The grand potential (or equivalently the effective action) has been obtained at one-loop order of the nucleon with the meson mean fields in Eq.~(\ref{OmegaPDM}), and the ``ground state'' of the system has been determined by solving the gap equations with respect to $\phi_N$, $\phi_S$, and $\bar{\omega}_N$ in Eq.~(\ref{GapEq}), respectively. 
Accordingly, the meson masses in nuclear matter are defined by including one-loop corrections in addition to the meson mean fields~\cite{Suenaga:2017deu,Suenaga:2017wbb}.

A self energy for the meson $X$ at one-loop order in momentum space generally depends on the external momentum, but here we consider $\Pi_X(q_0, \vec{q}=\vec{0})$. In our approach, since the one loops are regarded as corrections to the meson mean fields, we reduce the self energy to a local form approximately as $\Pi_X(q_0,\vec{0}) \to \Pi_X(m_X,\vec{0})$, with $m_X$ a mass of meson $X$ in the mean-field approximation defined in Eqs.~(\ref{S1}) -~(\ref{A1}). Because self energies of the mesons are intricate, detailed calculations of the one-loop self energies of the mesons are provided in Appendix~\ref{sec:Spin0SE}  for spin-$0$ mesons and in Appendix~\ref{sec:Spin1SE} for spin-$1$ mesons. If the meson $X$ is not affected by any mixings, then the medium mass of meson $X$ is simply given by $(m^2_X)^{\rm med} \equiv m_X^2 + \Pi_X(m_X,\vec{0})$. But in fact, some mesons mix with others as in Ref.~\cite{Parganlija:2012fy}, which forces us to solve them. We summarize the procedure to solve the mixings and each meson mass formula in medium in Appendix~\ref{sec:MassesInMedium}.

The resultant plots with several choices of $M_0$ and $\tilde{k}_1$ ($\tilde{k}_1 =k_1\hat{\phi}_N$) are depicted in Fig.~\ref{fig:M0800K10} - Fig.~\ref{fig:M0700K1-5}.\footnote{In these figures, $f_0^{L}$ and $f_0^H$ indicate the scalar-isoscalar mesons possessing smaller and larger masses in our model, respectively. $f_1^N$ and $f_1^S$ are the axial-vector--isoscalar mesons without and with the (anti-)strange quark, respectively. Also, $\omega_N$ and $\omega_S$ correspond to $\omega$ meson and $\phi$ meson in the Particle Data Group (PDG) notation.} Although the reproduction of nuclear matter properties allows the value of $\tilde{k}_1$ to be up to $\tilde{k}_1\approx 5$, when we take $\tilde{k}_1\approx 5$, the mass of $\eta$ meson drops significantly and becomes negative in the lower density regime ($1.5 \lesssim \rho/\rho_0$), caused by the contact interaction with the nucleon (described by the second term in Eq~(\ref{PiEtaN})). This behavior leads to an $\eta$ meson condensation phase that breaks parity, which should be discarded. In other words, the mass reduction of the $\eta$ meson is sensitive to the value of $\tilde{k}_1$. Experimentally, while the $\eta$-nucleus interaction is known to be attractive, its binding energy has not been determined well~\cite{Metag:2017yuh}. Our results suggest $\tilde{k}_1\approx 0$ is preferable so as to get an appropriate mass reduction at finite density.

Also, we find the mass of $f_0^L$ is $(m_{f_0^L})^{\rm vac} = 0.18\, {\rm GeV} - 0.27\, {\rm GeV}$ in vacuum, and increases as we access finite density. The Particle Data Group (PDG) shows the mass of the lightest scalar-isoscalar meson ($f_0(500)$) is in a range of $m_{f_0(500)} = 400$ - $500$ MeV~\cite{Tanabashi:2018oca}, which contradicts our results. However, the decay width of $f_0(500)$ is large as well: $\Gamma_{f_0(500)} = 400$ - $700$ MeV, such that we need to include the dynamical processes to estimate the $f_0(500)$ mass properly by the $f_0^L$ one in our model. We discuss this issue in Sec.~\ref{sec:Discussion} in detail.

%%%%%%%%%
\begin{figure*}[thbp]
\centering
\includegraphics*[scale=0.4]{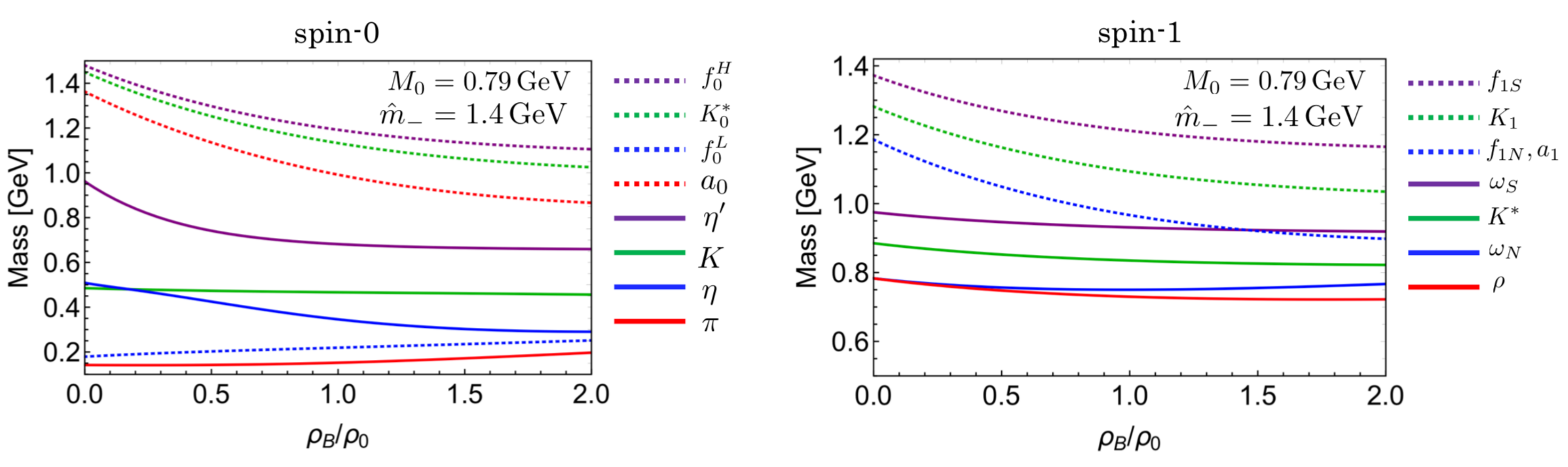}
\caption{(color online) The density dependence of spin-$0$ (left) and spin-$1$ (right) meson masses with $\hat{m}_-=1.4$ MeV, $M_0=0.79$ GeV, and $k_1=k_2=0$.}
\label{fig:M0790N-1400}
\end{figure*}
%%%%%%%%%
%%%%%%%%%
\begin{figure*}[thbp]
\centering
\includegraphics*[scale=0.4]{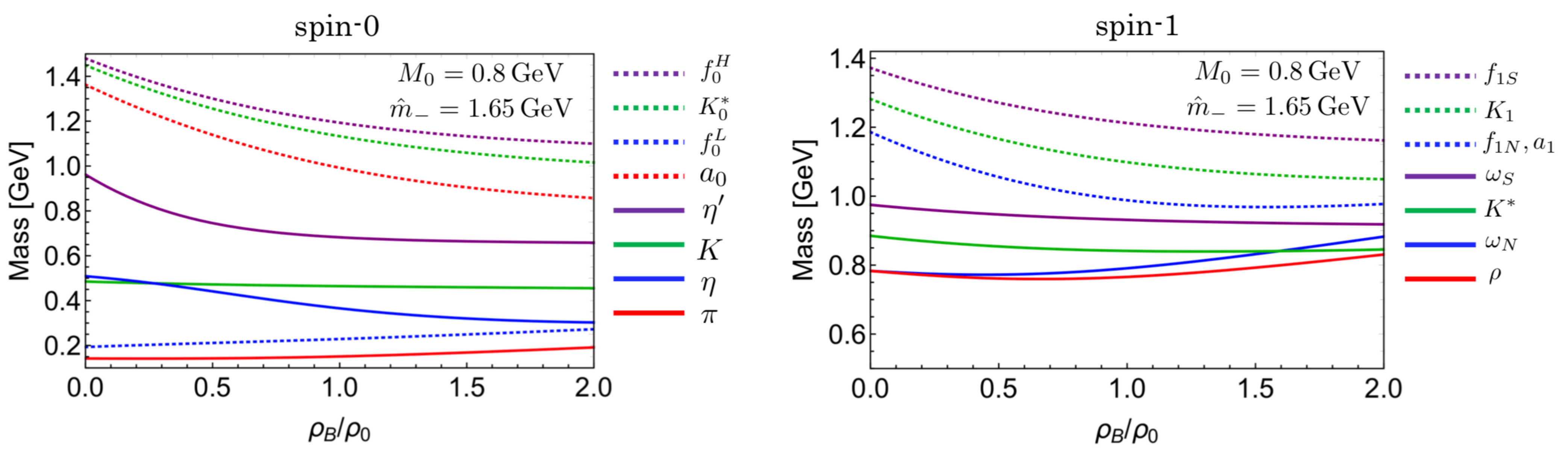}
\caption{(color online) The density dependence of spin-$0$ (left) and spin-$1$ (right) meson masses with $\hat{m}_-=1.65$ MeV, $M_0=0.8$ GeV, and $k_1=k_2=0$.}
\label{fig:M0800N-1650}
\end{figure*}
%%%%%%%%%

All figures indicate the masses of $f_0^H$, $K_0^*$, $a_0$, $\eta'$, $\eta$ mesons decrease at finite density. Especially, the mass reduction of $\eta'$ is about $200$ MeV at normal nuclear density $\rho_B=\rho_0$ for all cases, which is larger than those in previous studies~\cite{Costa:2002gk,Nagahiro:2006dr,Sakai:2013nba} where the estimations are about $100$ MeV. The large reduction of the $\eta'$ meson mass would support the possibility of the formation of $\eta'$ mesic nuclei as well as the effective restoration of $U(1)_A$ axial anomaly, but experimentally, the potential depth has been reported as $V=-(44 \pm 16({\rm stat}) \pm 15({\rm syst}))$ MeV~\cite{Nanova:2018zlz}. This discrepancy could be solved by employing a three-flavor PDM, which will be discussed in more detail in section \ref{sec:EtaImprove}.

Regarding the spin-$1$ mesons, the $f_{1S}$, $K_1$, $f_{1N}$, $a_1$, $\omega_S$, and $K^*$ meson masses decrease as the density increases for any choice of $M_0$ and $\tilde{k}_1$, whereas for the $\rho$ and $\omega_N$ meson masses it depends on the specific parameters. For $M_0=0.8$ MeV, the $\rho$ meson mass decreases slightly while the $\omega_N$ meson mass scarcely changes at $\rho_B=\rho_0$. On the other hand, for $M_0=0.7$ MeV, the $\rho$ meson mass does not change while the $\omega_N$ meson mass increases at $\rho_B=\rho_0$, and both masses increase at higher density. The experimental result shows the mass reduction of $\omega_N$ meson is $-(29\pm19({\rm stat})\pm20({\rm syst}))$ MeV at normal nuclear density whereas a large imaginary part of the optical potential of $70$ MeV is also expected~\cite{Metag:2017yuh}, such that the choice of $M_0=0.8$ GeV is preferable. In fact, we have confirmed that the smaller value of $M_0$ we take, the more rapidly the $\omega_N$ meson mass increases as we access the finite density. It is worth noting that the difference of the density dependence of $\rho$ and $\omega_N$ meson masses are induced by the difference between the $\rho N N$ and $\omega_N N N$ couplings which is allowed by the chiral symmetry as mentioned in Sec.~\ref{sec:PDM}.

The $\rho$ meson is regarded as chiral partner to the $a_1$ meson within the two-flavor chiral symmetry such that the $\rho$ meson mass shift in nuclear matter is significant to study the partial restoration of chiral symmetry in medium~\cite{Hatsuda:1991ez,Rapp:1999ej}. Although all figures show that the $\rho$ and $a_1$ mesons tend to degenerate at higher density, we need to include the decay widths and a broadening effect in order to study the mass shifts more precisely~\cite{Chanfray:1993ue,Brown:1998ca,Rapp:1999us}.

The mass reduction of $f_{1N}$ meson at normal nuclear density is about $150$ - $200$ MeV, which is larger than the result in Ref.~\cite{Gubler:2016djf} using the QCD sum rule approach in which the mass reduction is estimated from $55\, {\rm MeV}$ to $130\, {\rm MeV}$. In the above reference, the  $\omega$ and $f_{1N}$ mesons are regarded as chiral partners in the context of two-flavor chiral symmetry within the large-$N_c$ limit. In our calculation, although the chiral symmetry is not restored sufficiently, we can clearly observe a tendency of degeneracy of the $\omega_N$ and $f_{1N}$ mesons. In terms of the $\omega_S$ meson, we find a mass reduction of a few \%, which is comparably consistent with the experimental data~\cite{Muto:2005za}. This small mass reduction is realized within our model by assuming the large-$N_c$ suppression, {\it i.e.}, by dropping $h_1$ term in Eq.~(\ref{eLSMOriginal}).

As mentioned earlier, for the fitting procedure yielding Table~\ref{tab:ParametersPD} we have assumed that the axial charge of $N^*(1535)$ is $\hat{g}_A^{N_-} = 0.2$ while the value includes a large uncertainty as indicated in Table~\ref{tab:InputVac}. Therefore, we have varied the value from $\hat{g}_A^{N_-} = -0.1$ to $\hat{g}_A^{N_-}=0.5$ and found that all masses are insensitive to the value of $\hat{g}_A^{N_-}$.

\subsection{Other choices for $N_-$}
\label{sec:OtherM-}

While we have assigned the $N^*(1535)$ to $N_-$ in Sec.~\ref{sec:NResults}, it is possible to regard another nucleon as the chiral partner to the nucleon $N(939)$. To examine such a possibility, we simply change the input parameter $\hat{m}_-$ to $\hat{m}=1.2$ GeV, $\hat{m}_-=1.4$ GeV and $\hat{m}_-=1.65$ GeV whereas the other inputs except for the $N_-\to N_+\eta$ decay width are unchanged, since the mass modification of the mesons has been found to be less sensitive to $g_V$ and $h_V$. The values of $k_1$ and $k_2$ are fixed to be zero here for simplicity.

%%%%%%%%%%%%%%%%%%%%%%%%%
\begin{table}[thbp]
  \begin{tabular}{c|c} \hline\hline 
$\hat{m}_-$ [GeV]     &       Range of $M_0$ [GeV] \\\hline
$1200$ & 0.57 - 0.75 \\
$1400$ & 0.60  - 0.79 \\
$1535$ & 0.61  - 0.81 \\
$1650$ & 0.62  - 0.82  \\ \hline \hline
  \end{tabular}
\caption{The allowed value of $M_0$ for each $\hat{m}_-$.}
  \label{tab:AllowedM0}
\end{table}
%%%%%%%%%%%%%%%%%%%%%%%%%
%%%%%%%%%%%%%%%%%%%%%%%%%
\begin{table*}[thbp]
  \begin{tabular}{c||cccccccccc} \hline\hline 
$\hat{m}$ [GeV] & $M_0$ [GeV] &  ${k}_1$ & $G_1$ & $G_2$  & ${k}_2$ & $g_V$ & $h_V$ & $\tilde{g}$ &  $\lambda_1$ & $g_{4p}$ \\\hline
1.4 & 0.79 & 0 & 3.839 & 6.639 & 0 & 4.060 & -8.973 & -2.531 & -23.01 & 3.420 \\ \hline
1.65 & 0.8 & 0 & 4.023 & 8.343 & 0 & 3.661 & -9.372 & -2.792 & -22.95 & 89.84 \\ \hline \hline
  \end{tabular}
\caption{
Parameters for $\hat{m}_- = 1.4$ GeV and $\hat{m}_- = 1.65$ GeV for the plots in Fig.~\ref{fig:M0790N-1400} and Fig.~\ref{fig:M0800N-1650}. }
  \label{tab:PM-Change}
\end{table*}
%%%%%%%%%%%%%%%%%%%%%%%%%
A range of allowed values of $M_0$ for each choice of $\hat{m}_-$ is listed in Table~\ref{tab:AllowedM0}. This table shows the smaller value of $\hat{m}_1$ we take, the smaller value of $M_0$ we can obtain, as can be anticipated naively. We also plot the resultant density dependence of meson masses in Fig.~\ref{fig:M0790N-1400} with $\hat{m}_-=1.4$ MeV, $M_0=0.79$ GeV, and in Fig.~\ref{fig:M0800N-1650} with $\hat{m}_-=1.65$ MeV, $M_0=0.8$ GeV. The other parameters are listed in Table~\ref{tab:PM-Change}. Fig.~\ref{fig:M0790N-1400} shows a rather reasonable value of $\omega_N$ meson mass at $\rho_0$. When we plot the result with $\hat{m}=1.2$ GeV, the $\eta$ mass turns into imaginary below $\rho_B \approx1.3\rho_0$ which is unphysical for any allowed values of $M_0$.

%%%%%%%%%%%%%%%%%%%%%

%%%%%%%%%
\begin{figure}[thbp]
\centering
\includegraphics*[scale=0.26]{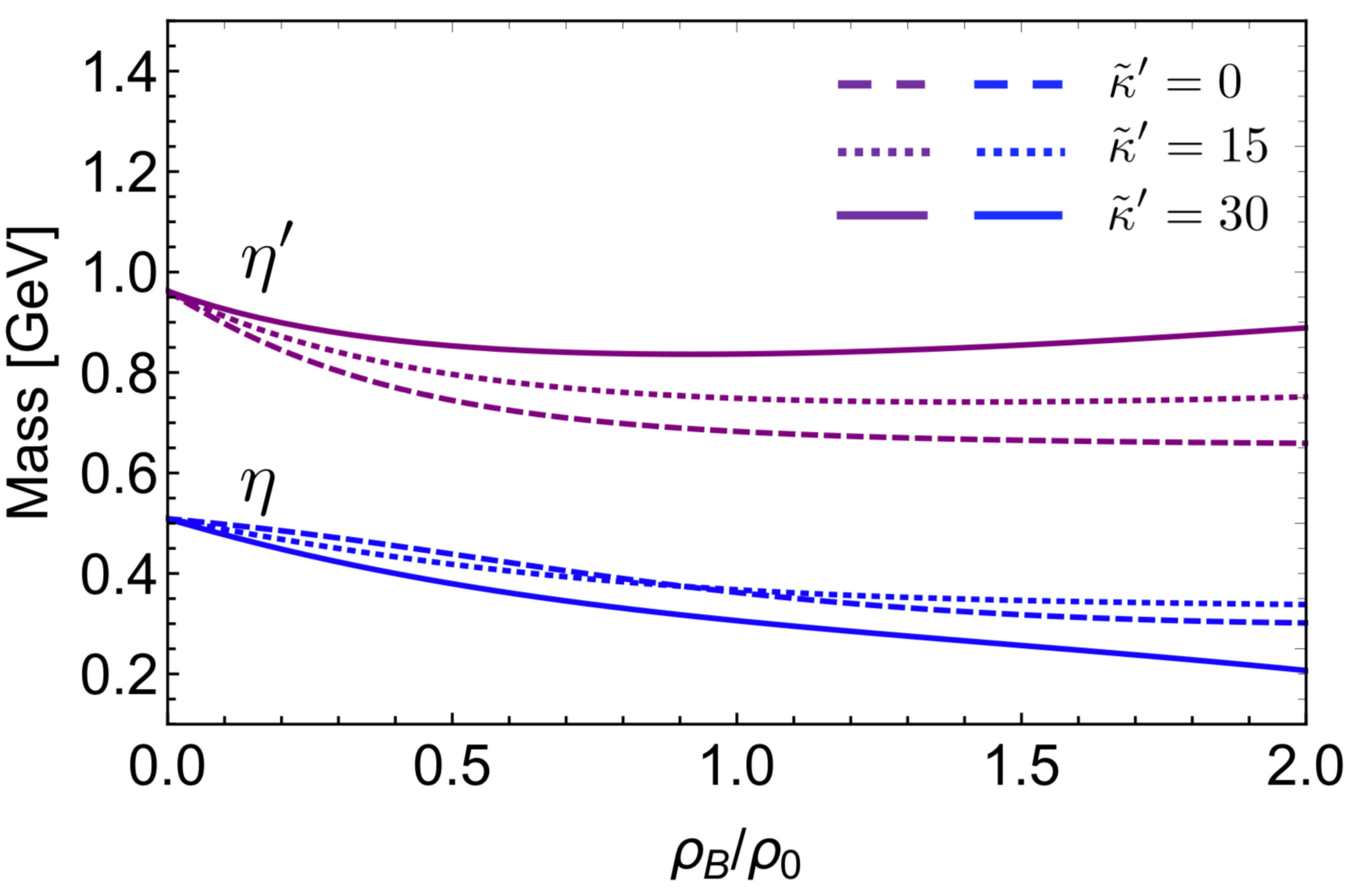}
\caption{(color online) The density dependence of $\eta'$ (purple) and $\eta$ (blue) masses with the interaction term~(\ref{ThreeAnomaly}).  The dashed, dotted and solid curves correspond to $\tilde{\kappa}'=0$, $\tilde{\kappa}'=15$ and $\tilde{\kappa}'=30$, respectively.}
\label{fig:EtaEtaPrime}
\end{figure}
%%%%%%%%%

\section{Discussions}
\label{sec:EtaImprove}
In this section we discuss the density dependences of $\eta$ and $\eta'$ mesons. The results in Sec.~\ref{sec:Results} show that the $\eta'$ meson mass is reduced by approximately 200 MeV at normal nuclear density $\rho_B=\rho_0$, which is greatly larger than the experimental result of $V=-(44 \pm 16({\rm stat}) \pm 15({\rm syst}))$ MeV~\cite{Nanova:2018zlz}. This discrepancy could be improved by extending the two-flavor PDM to the three-flavor PDM. 

For instance, following the procedure in Ref.~\cite{Olbrich:2017fsd,Olbrich:2015gln}, the $k_1$ term describing $U(1)_A$ anomaly in Eq.~(\ref{LNStart}) is replaced by
\begin{eqnarray}
{\cal L}^{\rm 3f-PDM}_{\rm anomaly}\Big|_{\eta_N\eta_S} = -\frac{\kappa'}{2\sqrt{2}}(2\phi_N\eta_N\eta_S+\phi_S\eta_N^2)\bar{N}_+N_+\ ,  \nonumber\\ \label{ThreeAnomaly}
\end{eqnarray}
(the factor $1/(2\sqrt{2})$ is not important) when we employ the three-flavor version of PDM. Here, to demonstrate how the $\eta'$ meson mass in nuclear matter is improved by Eq.~(\ref{ThreeAnomaly}), we only replace the $k_1$ term in Eq.~(\ref{LNStart}) by Eq.~(\ref{ThreeAnomaly}), and leave all other terms unchanged. Besides, we adopt the parameter set of the second line in Table~\ref{tab:ParametersPD} apart from $\tilde{k}_1$. The resultant density dependence of $\eta$ and $\eta'$ meson masses with $\tilde{\kappa}' = 0,15,30$ ($\tilde{\kappa}'={\kappa}'\hat{\phi}_N^2$) is depicted in Fig.~\ref{fig:EtaEtaPrime}. The dashed, dotted and solid curves correspond to the results with $\tilde{\kappa}'=0$, $\tilde{\kappa}'=15$ and $\tilde{\kappa}'=30$, respectively. The figure shows that when we use the larger value for $\tilde{\kappa}'$, the resultant mass reduction of $\eta'$ meson in nuclear matter gets small. Note that the result with $\tilde{\kappa}'=0$ coincides with the one in Fig.~\ref{fig:M0800K10}.

The demonstration provided above implies that the large mass reduction of $\eta'$ meson in nuclear matter obtained in Sec.~\ref{sec:Results} can be improved by employing the three-flavor PDM, especially by the $U(1)_A$ breaking term. In the same way, the $\kappa'$ term also generates ${\cal L}^{\rm 3f-PDM}_{\rm anomaly}|_{K} =\kappa' \frac{\phi_N}{2}K^+K^-\bar{N}_+N_+ $ and the resultant mass reduction of the kaon at $\rho_B=\rho_0$ reads approximately $100$ MeV for $\tilde{\kappa}'=15$ and $230$ MeV for $\tilde{\kappa}'=30$, respectively. These results show that the kaon mass also becomes dependent on the density when we employ the three-flavor PDM, while experimentally a mass shift of the $K^-$ meson to about 270 MeV at $\rho_B=2\rho_0$ has been observed~\cite{Schroter:1994ck,Barth:1997mk}(for a mass shift of the $K^+$ meson, see, {\it e.g.}~\cite{Hartnack:2011cn}).\footnote{The mass reduction of the $K^-$ meson by about 100 MeV at $\rho_B=\rho_0$ is found in the Quark Meson Coupling model, see, {\it e.g.}~Ref.\cite{Saito:2005rv}.}

The treatment employed in this section violates the chiral symmetry explicitly because we have assumed that $\kappa'$ does not depend on density, although $\kappa'$ is a function of the density-dependent nucleon mass. Besides, for more precise understanding within the three-flavor PDM, we need to take other interacting terms allowed by chiral symmetry into account in addition to Eq.~(\ref{ThreeAnomaly})~\cite{Nishihara:2015fka,Olbrich:2015gln}. Such a global analysis is beyond the scope of present work and left for a future publication. The mass splitting between $K$ and $\bar{K}$ can also be generated within such a global analysis. However, the density dependence of spin-$1$ meson masses is expected to be largely unchanged compared with the current study even if we employ the three-flavor PDM, since the interactions among spin-$1$ mesons and nucleons in the three-flavor PDM are essentially the same as the ones in Eq.~(\ref{LNStart})~\cite{Olbrich:2015gln}.

%%%%%%%%%%%%%%%%%%%%%

\section{Conclusions}
\label{sec:Discussion}
In this study we have investigated the mass of scalar, pseudo-scalar, vector, and axial-vector mesons in nuclear matter comprehensively by employing the three-flavor extended Linear Sigma Model and the two-flavor Parity Doublet Model. To this end, we have calculated one-loop corrections by the nucleons to the meson mean fields. To fix the model parameters, vacuum properties of the nucleons as well as normal nuclear matter properties are used as inputs as shown in Table~\ref{tab:InputVac} and Table~\ref{tab:InputMedium}. Due to a strong restriction by the latter inputs, we find the value of the chiral invariant mass ($M_0$) should be in a range of $0.6\, {\rm GeV}\lesssim M_0\lesssim 0.8\, {\rm GeV}$, when we regard $N^*(1535)$ as the chiral partner to the nucleon.

The results show that all spin-$0$ meson masses except the $\pi$, $K$, and the lightest scalar-isoscalar ($f_0^L$) ones decrease at finite baryon density. Especially, a mass reduction of $\eta'$ meson is about $200$ MeV, which is larger than the previous works~\cite{Costa:2002gk,Nagahiro:2006dr,Sakai:2013nba}. Also, in terms of the direct $U(1)_A$ axial anomaly contribution to the nucleons, we find $k_1\approx0$ ($k_1$ is given in Eq.~(\ref{LNStart})) so as to obtain an appropriate mass reduction of the $\eta$ meson at normal nuclear density.

For spin-$1$ mesons, all axial-vector meson masses decrease at finite density, while density dependences of $\rho$ and $\omega_N$ mesons depend on the value of $M_0$. Especially, the $\omega_N$ meson mass increases at finite density when we take a smaller value of $M_0$. The experimental result suggests a small reduction of the $\omega_N$ meson mass at normal nuclear density~\cite{Metag:2017yuh}, hence, to reproduce such behavior, $M_0 \approx 0.8$ GeV is preferable within our framework. Unlike for spin-$0$ mesons, the large-$N_c$ suppression is assumed in our approach so that the chiral partner structure of the $\omega_N$ and $f_{1N}$ can be observed as proposed in~\cite{Gubler:2016djf}.

We expect our results to provide useful information on meson mass shifts in nuclear matter to existing and upcoming experiments, especially with regard to the partial restoration of chiral symmetry and $U(1)_A$ axial anomaly restoration.

In what follows, we discuss topics which are not covered in this paper. The small vacuum mass of the $f_{0}^{L}$ and the large mixing with $f_{0}^{H}$ might be an indication that another scalar-isoscalar resonance is needed to obtain correct vacuum values for the $f_0$ mesons. 
For instance, in Ref.~\cite{Gallas:2011qp} the authors found that at nonzero density a light tetraquark has a strong influence on the medium properties of the system due to the interplay of two condensates, the tetraquark and the chiral condensate.
A tetraquark degree of freedom $\chi$ in the leading order of the large-$N_c$ expansion in the two-flavor case can be incorporated into the current model using the following interaction terms \cite{Lakaschus:2018rki}:
\begin{eqnarray}
\mathcal{L}_{\chi \Phi \Phi} &= \dfrac{c}{2} \chi \left(\sigma_N^2 + \vec{\pi}^2 - \vec{a_0}^2 - \eta_N^2\right) \, , \\
\mathcal{L}_{\chi AV} &= \dfrac{d}{2} \chi \left(\vec{\rho}_\mu^2 + \vec{a}_{1,\mu}^2 - \omega_\mu^2 - f_{1, \mu}^2 \right) \, ,
\end{eqnarray}
where the new eLSM Lagrangian would be given as:
\begin{eqnarray}
\mathcal{L}_{\textrm{eLSM}} \rightarrow \mathcal{L}_{\textrm{eLSM}}  + \dfrac{1}{2} \partial_\mu \chi \partial^\mu \chi - \dfrac{m_\chi^2}{2} \chi^2 + \mathcal{L}_{\chi \Phi \Phi} +  \mathcal{L}_{\chi AV} \, . \nonumber\\ \label{TetraQ}
\end{eqnarray}
In this framework the tetraquark is assumed to be mostly $f_0(500)$, while the $\sigma_N$ and $\sigma_S$ correspond most likely to $f_0(1370)$ and $f_0(1700)$, respectively, as previously eLSM studies have shown \cite{Lakaschus:2018rki, Janowski:2011gt, Parganlija:2010fz}. Note, while in Ref. \cite{Lakaschus:2018rki} the authors found a negligible tetraquark condensate, the situation here might be different due to the inclusion of the $\phi_S$ condensate which is not considered in Ref. \cite{Lakaschus:2018rki}. Modification of our model by including the tetraquark as in Eq.~(\ref{TetraQ}) will be left as a future work.

%%%%%%%%%%%%%%%%%%%%%

\acknowledgments

D.S. is supported by NSFC grant 20201191997, and wishes to thank Johann Wolfgang Goethe University for its hospitality during his stay there. The authors thank F. Giacosa, P. Gubler, M. Harada, T. Hyodo, D. H. Rischke, S. Sakai and J. Schaffner-Bielich for useful discussions and comments. We also thank V. Metag for useful comments and information on recent experimental results.

%\newpage

\appendix

\section{Meson masses in vacuum}
\label{sec:MeanMass}
Here, we show mass formulae of scalar, pseudo-scalar, vector, and axial-vector mesons in vacuum obtained by the three-flavor eLSM in Eq.~(\ref{eLSM}). By reading the quadratic terms with respect to the meson fields in Eq.~(\ref{eLSM}) in the presence of the mean fields in Eq.~(\ref{MeanField}), first we find 
\begin{widetext}
\begin{eqnarray}
{\cal L}_{\rm eLSM}^{{\rm red}}&=& {\cal L}^{(2)}_{\sigma \omega} + {\cal L}^{(2)}_{a_0\rho} + {\cal L}^{(2)}_{K_0^* K^*} + {\cal L}^{(2)}_{\eta f_1} + {\cal L}^{(2)}_{\pi a_1} + {\cal L}^{(2)}_{KK_1}   + \cdots\ ,
 \end{eqnarray}
with
\begin{eqnarray}
{\cal L}^{(2)}_{\sigma \omega} &=& \frac{1}{2}\partial_\mu\sigma_N\partial^\mu\sigma_N-\frac{m_{\sigma_N}^2}{2}\sigma_N^2 + \frac{1}{2}\partial_\mu\sigma_S\partial^\mu\sigma_S-\frac{m_{\sigma_S}^2}{2}\sigma_S^2-m^2_{\sigma_N\sigma_S}\sigma_N\sigma_S \nonumber\\
&&  -\frac{1}{4}\omega_{N\mu\nu}\omega_N^{\mu\nu}  +\frac{m_{\omega_N}^2}{2}\omega_{N\mu}\omega_N^\mu + 2g_{4p}\bar{\omega}_N^2(\omega_N^{\mu=0})^2  -\frac{1}{4}\omega_{S\mu\nu}\omega_S^{\mu\nu}  +\frac{m_{\omega_S}^2}{2}\omega_{S\mu}\omega_S^\mu \ ,
\end{eqnarray}
\begin{eqnarray}
{\cal L}_{a_0\rho}^{(2)} = \frac{1}{2}\partial_\mu a_0^a\partial^\mu a_0^a-\frac{m_{a_0}^2}{2}a_0^aa_0^a  -\frac{1}{4}\rho_{\mu\nu}^a\rho^{ a\mu\nu}+\frac{m_\rho^2}{2}\rho_{\mu}^a\rho^{a\mu} + 2g_{4p}\bar{\omega}_N^2(\rho^{\mu=0})^2\ , \label{LA0Rho}
\end{eqnarray}
\begin{eqnarray}
{\cal L}^{(2)}_{K_0^*K^*}  &=& \partial_\mu \bar{K}_0^*\partial^\mu K_0^*+ig_1\bar{\omega}_N\partial_0 K_0^*\bar{K}_0^*-m_{K_0^*}^2\bar{K}_0^*K_0^* + ig_1\left(\frac{\phi_S}{\sqrt{2}}-\frac{\phi_N}{2}\right)(\bar{K}_\mu^{*}\partial^\mu K_0^{*}-\partial^\mu\bar{K}_0^{*}K_\mu^{*})  \nonumber\\
&&-\frac{1}{2}\bar{K}_{\mu\nu}^*K^{*\mu\nu}+m_{K^*}^2\bar{K}_\mu^{*}K^{*\mu}+ \frac{1}{2}g_{4p}\bar{\omega}_N^2 \bar{K}^{*\mu=0} K^{*\mu=0} \ , \label{LK0stKst}
\end{eqnarray}
\begin{eqnarray}
{\cal L}_{\eta f_1}^{(2)} &=& \frac{1}{2}\partial_\mu\eta_N\partial^\mu\eta_N-\frac{m_{\eta_N}^2}{2}\eta_N^2 + \frac{1}{2}\partial_\mu\eta_S\partial^\mu\eta_S-\frac{m_{\eta_S}^2}{2}\eta_S^2-m^2_{\eta_N\eta_S}\eta_N\eta_S-g_1\phi_N\partial_\mu\eta_N f_{1N}^\mu  -\sqrt{2}g_1\phi_S\partial_\mu\eta_S f_{1S}^\mu \nonumber\\
&& -\frac{1}{4}f_{1N\mu\nu}f_{1N}^{\mu\nu}+\frac{m_{f_{1N}}^2}{2}f_{1N\mu} f_{1N}^\mu -\frac{1}{4}f_{1S\mu\nu}f_{1S}^{\mu\nu}+\frac{m_{f_{1S}}^2}{2}f_{1S\mu} f_{1S}^\mu+ 2g_{4p}\bar{\omega}_N^2(f_{1N}^{\mu=0})^2\ , \label{LEtaF1}
\end{eqnarray}
\begin{eqnarray}
{\cal L}_{\pi a_1}^{(2)} =  \frac{1}{2}\partial_\mu \pi^a\partial^\mu \pi^a-\frac{m_{\pi}^2}{2}\pi^a\pi^a-g_1\phi_N \partial_\mu\pi^aa_1^{a\mu}  -\frac{1}{4}a_{1\mu\nu}^aa_1^{a\mu\nu}+\frac{m_{a_1}^2}{2}a_{1\mu}^a a_1^{a\mu} + 2g_{4p}\bar{\omega}_N^2(a_1^{\mu=0})^2\ , \label{LPiA1}
\end{eqnarray}
\begin{eqnarray}
{\cal L}^{(2)}_{KK_1} &=& \partial_\mu\bar{K}\partial^\mu K+ig_1\bar{\omega}_N\partial_0 K\bar{K}-m_K^2\bar{K}K -g_1\left(\frac{\phi_S}{\sqrt{2}}+\frac{\phi_N}{2}\right)(\bar{K}_{1\mu}\partial^\mu K+\partial^\mu\bar{K}K_{1\mu}) \nonumber\\
&&-\frac{1}{2}\bar{K}_{1\mu\nu}K_1^{\mu\nu}+m_{K_1}^2\bar{K}_{1\mu}K_1^{\mu} + \frac{1}{2}g_{4p}\bar{\omega}_N^2\bar{K}_1^{\mu=0} K_1^{\mu=0}\ , \label{LKK1}
\end{eqnarray}
where we have defined 
\begin{eqnarray}
{m}_{\sigma_N}^{2} &=& m_0^2+\left(3\lambda_1+\frac{3}{2}\lambda_2\right)\phi_N^2+\lambda_1\phi_S^2- \frac{1}{2}(h_2+h_3) \bar{\omega}_N^2 \ , \nonumber\\
m_{\sigma_S}^{2} &=&m_0^2+\lambda_1\phi_N^2+3(\lambda_1+\lambda_2)\phi_S^2  \ ,\nonumber\\
m_{\sigma_S\sigma_N}^{2}&=& 2\lambda_1\phi_N\phi_S \ ,\nonumber\\
{m}_{a_0}^{2} &=& m_0^2+\left(\lambda_1+\frac{3}{2}\lambda_2\right)\phi_N^2+\lambda_1\phi_S^2 - \frac{1}{2}(h_2+h_3) \bar{\omega}_N^2 \nonumber\\
m_{K_0^*}^2 &=& m_0^2+\left(\lambda_1+\frac{\lambda_2}{2}\right)\phi_N^2+(\lambda_1+\lambda_2)\phi_S^2+\frac{\lambda_2}{\sqrt{2}}\phi_N\phi_S-g_1^2\frac{\bar{\omega}_N^2}{4}-\frac{1}{4}(h_2+h_3)\bar{\omega}_N^2  \ , \label{S1}
\end{eqnarray}
for scalar mesons,
\begin{eqnarray}
{m}_{\eta_N}^2 &=& m_0^2+\left(\lambda_1+\frac{\lambda_2}{2}\right)\phi_N^2+\lambda_1\phi_S^2+c_1\phi_N^2\phi_S^2- \frac{1}{2}(h_2+h_3) \bar{\omega}_N^2\ ,\nonumber\\
m_{\eta_S}^{2} &=& m_0^2+\lambda_1\phi_N^2+(\lambda_1+\lambda_2)\phi_S^2+\frac{c_1}{4}\phi_N^4\ , \nonumber\\
m^{2}_{\eta_N\eta_S} &=& \frac{c_1}{2}\phi_N^3\phi_S\ , \nonumber\\
{m}_\pi^2 &=& m_0^2+\left(\lambda_1+\frac{\lambda_2}{2}\right)\phi_N^2+\lambda_1\phi_S^2 -\frac{1}{2}(h_2+h_3) \bar{\omega}_N^2 \ ,\nonumber\\
m_K^2 &=&m_0^2+\left(\lambda_1+\frac{\lambda_2}{2}\right)\phi_N^2+(\lambda_1+\lambda_2)\phi_S^2-\frac{\lambda_2}{\sqrt{2}}\phi_N\phi_S -g_1^2\frac{\bar{\omega}_N^2}{4}- \frac{1}{4}(h_2+h_3) \bar{\omega}_N^2 \ , \label{P1}
\end{eqnarray}
for pseudo-scalar mesons,
\begin{eqnarray}
{m}_{\omega_N}^{2} &=& m_1^2+\frac{1}{2}(h_2+h_3)\phi_N^2+ 2g_{4p}\bar{\omega}_N^2\ , \nonumber\\
m_{\omega_S}^2 &=& m_1^2+\left(h_2+h_3\right)\phi_S^2+2\delta_S \ ,\nonumber\\
m_\rho^2 &=& m_{\omega_N}^2\ , \nonumber\\
m_{K^*}^2 &=& m_1^2+\frac{1}{4}(g_1^2+h_2)\phi_N^2+\frac{1}{2}(g_1^2+h_2)\phi_S^2+\frac{1}{\sqrt{2}}(h_3-g_1^2)\phi_N\phi_S +\delta_S  + g_{4p}\bar{\omega}_N^2\ , \label{V1}
\end{eqnarray}
for vector-mesons, and
\begin{eqnarray}
{m}_{f_{1N}}^{2} &=& m_1^2+\frac{1}{2}(2g_1^2+h_2-h_3)\phi_N^2 +  2g_{4p}\bar{\omega}_N^2 \ , \nonumber\\
m_{f_{1S}}^2 &=& m_1^2+(2g_1^2+h_2-h_3)\phi_S^2+2\delta_S\ , \nonumber\\
{m}_{a_{1}}^{2}  &=& {m}_{f_{1N}}^{2}\ ,  \nonumber\\
m_{K_1}^2 &=&m_1^2+\frac{1}{4}(g_1^2+h_2)\phi_N^2+\frac{1}{2}(g_1^2+h_2)\phi_S^2-\frac{1}{\sqrt{2}}(h_3-g_1^2)\phi_N\phi_S+\delta_S + g_{4p}\bar{\omega}_N^2 \ ,\label{A1}
\end{eqnarray}
\end{widetext}
for axial-vector mesons, respectively. With respect to $K$ ($\bar{K}$) and $K_0^*$ $(\bar{K}_0^*)$ mesons, we have defined the mass at finite chemical potential in the same manner as the ``effective mass'' in Ref.~\cite{Schaffner:1996kv}~\footnote{If we define the mass by $m_{\cal K} \equiv \omega_{\cal K}(|\vec{k}|=0)$ (${\cal K} = K, \bar{K}, K_0^*,\bar{K}_0^*$) with $\omega_{\cal K}$ the dispersion, then the masses of $K$ and $\bar{K}$ ($K_0^*$ and $\bar{K}_0^*$) split.}.

In Eqs.~(\ref{LA0Rho}) -~(\ref{LKK1}), a term proportional to $|{\cal V}^{\mu=0}|^2$ (${\cal V} = \rho,K^*,f_{1N},a_1,K_1$) is present. Field theoretically, one simple way to remove the unphysical mode of the massive spin-$1$ meson is to start on a perturbation series by a Proca-type Lagrangian. Thus, in order to restrict ourselves to the Proca-type Lagrangian, we simply discard such a problematic term.

The masses of $a_0$, $\omega_N$, $\omega_S$, $\rho$, $K^*$, $f_{1N}$, $f_{1S}$, $a_1$, $K_1$ in vacuum are straightforwardly obtained as
\begin{eqnarray}
&&(m_{a_0}^2)^{\rm vac} = \hat{m}_{a_0}^2\, ,  \ (m_{\omega_N}^2)^{\rm vac} = \hat{m}_{\omega_N}^2\, , \ (m_{\omega_S}^2)^{\rm vac} = \hat{m}_{\omega_S}^2 \, , \nonumber\\
&& (m_\rho^2)^{\rm vac} = \hat{m}_\rho^2\, ,\ (m_{K^*}^2)^{\rm vac} = \hat{m}_{K^*}^2\, ,  \ (m_{f_{1N}}^2)^{\rm vac}=\hat{m}_{f_{1N}}^2\, , \nonumber\\
&& (m_{f_{1S}}^2)^{\rm vac}=\hat{m}_{f_{1S}}^2 \, , \ (m_{a_1}^2)^{\rm vac}= \hat{m}_{a_1}^2\, , \ (m_{K_1}^2)^{\rm vac}= \hat{m}_{K_1}^2\, , \nonumber\\ \label{MassesNaive}
\end{eqnarray} 
where $\hat{m}_X^2$ ($X=\sigma_N,\sigma_S,a_0,\cdots$) represents the corresponding mean-field masses in Eqs.~(\ref{S1})-(\ref{A1}) in which $\phi_N$, $\phi_S$ and $\bar{\omega}_N$ are replaced by $\hat{\phi}_N$, $\hat{\phi}_S$ and $0$, respectively.
For the other mesons, we need to solve the mixings. As done in Ref.~\cite{Parganlija:2012fy}, by introducing mixing angles and redefining the spin-$1$ meson fields appropriately, we find
\begin{eqnarray}
&&(m_\pi^2)^{\rm vac} = \hat{Z}_\pi^2\hat{m}_\pi^2 \ , \ (m_{K_0^*}^2)^{\rm vac} = \hat{Z}_{K_0^*}^2\hat{m}_{K_0^*}^2 \ , \nonumber\\
&&(m_K^2)^{\rm vac} = \hat{Z}_K^2\hat{m}_K^2 \ , \label{MassesRenorm}
\end{eqnarray}
with
\begin{eqnarray}
\hat{Z}_{\pi} &=& \frac{\hat{m}_{a_1}}{\sqrt{\hat{m}_{a_1}^2-g_1^2\hat{\phi}_N^2}} \ ,\nonumber\\
\hat{Z}_{K_0^*} &=& \frac{2\hat{m}_{K^*}}{\sqrt{4\hat{m}_{K^*}^2-g_1^2(\hat{\phi}_N-\sqrt{2}\hat{\phi}_S)^2}} \ ,\nonumber\\
\hat{Z}_{K} &=& \frac{2\hat{m}_{K_1}}{\sqrt{4\hat{m}_{K_1}^2-g_1^2(\hat{\phi}_N+\sqrt{2}\hat{\phi}_S)^2}} \ , \label{ZFactorsVac}
\end{eqnarray}
for $\pi$, $K_0^*$ and $K$, while
\begin{eqnarray}
&&({m}_{f_0^H/f_0^L}^2)^{\rm vac} = \frac{1}{2}\Big(\hat{m}_{\sigma_N}^2+\hat{m}_{\sigma_S}^2 \nonumber\\
&&\ \ \ \ \ \ \ \ \ \ \ \  \ \ \ \ \ \ \ \  \pm\sqrt{(\hat{m}_{\sigma_N}^2-\hat{m}_{\sigma_S}^2)^2+4\hat{m}_{\sigma_N\sigma_S}^4}\Big) \ ,\nonumber\\
&& ({m}_{\eta'/\eta}^2)^{\rm vac} = \frac{1}{2} \Big(({m}_{\eta_N}^2)^{\rm vac}+({m}_{\eta_S}^2)^{\rm vac}  \nonumber\\
&& \ \ \ \ \pm \sqrt{\big(({m}_{\eta_N}^2)^{\rm vac}-({m}_{\eta_S}^2)^{\rm vac} \big)^2+4({m}_{\eta_N\eta_S}^4)^{\rm vac}}\Big)\ ,\nonumber\\ \label{MassF0Eta}
\end{eqnarray}
in which $(m_{\eta_N}^2)^{\rm vac} = \hat{Z}_{\eta_N}^2\hat{m}_{\eta_N}^2$, $(m_{\eta_S}^2)^{\rm vac} = \hat{Z}_{\eta_S}^2\hat{m}_{\eta_S}^2$ and $(m_{\eta_N\eta_S}^4)^{\rm vac} = \hat{Z}_{\eta_N}^2\hat{Z}_{\eta_S}^2\hat{m}_{\eta_N\eta_S}^4$ with
\begin{eqnarray}
\hat{Z}_{\eta_N} &=& \frac{\hat{m}_{f_{1N}}}{\sqrt{\hat{m}_{f_{1N}}^2-g_1^2\hat{\phi}_N^2}} \ ,\nonumber\\
\hat{Z}_{\eta_S} &=& \frac{\hat{m}_{f_{1S}}}{\sqrt{\hat{m}^2_{f_{1S}}-2g_1^2\hat{\phi}_S^2}} \ , \label{ZEtaVac}
\end{eqnarray}
for $\sigma_N$, $\sigma_S$, $\eta_N$, and $\eta_S$. In obtaining Eq.~(\ref{MassF0Eta}), we have diagonalized the mass matrices as
\begin{eqnarray}
\left(
\begin{array}{c}
f_0^L \\
f_0^H \\
\end{array}
\right) = \left(
\begin{array}{cc}
{\rm cos}\, \hat{\theta}_\sigma & -{\rm sin}\, \hat{\theta}_\sigma \\
{\rm sin}\, \hat{\theta}_\sigma & {\rm cos}\, \hat{\theta}_\sigma \\
\end{array}
\right)\left(
\begin{array}{c}
\sigma_N \\
\sigma_S \\
\end{array}
\right)\ ,
\end{eqnarray}
and
\begin{eqnarray}
\left(
\begin{array}{c}
\eta \\
\eta' \\
\end{array}
\right) = \left(
\begin{array}{cc}
{\rm cos}\, \hat{\theta}_\eta &- {\rm sin}\, \hat{\theta}_\eta \\
{\rm sin}\, \hat{\theta}_\eta & {\rm cos}\, \hat{\theta}_\eta \\
\end{array}
\right)\left(
\begin{array}{c}
\eta_N \\
\eta_S \\
\end{array}
\right)\ , \label{EtaMix}
\end{eqnarray}
by introducing mixing angles $\hat{\theta}_\sigma$ and $\hat{\theta}_\eta$ satisfying
\begin{eqnarray}
{\rm tan}\, 2\hat{\theta}_\sigma = \frac{2\hat{m}_{\sigma_N\sigma_S}^2}{\hat{m}_{\sigma_S}^2-\hat{m}_{\sigma_N}^2}\ ,\ \ {\rm tan}\, 2\hat{\theta}_\eta = \frac{2\hat{m}_{\eta_N\eta_S}^2}{\hat{m}_{\eta_S}^2-\hat{m}_{\eta_N}^2} \ .\nonumber\\
\end{eqnarray}

\section{Meson masses in nuclear matter}
\label{sec:MassesInMedium}

In this appendix, we discuss general properties of meson masses in nuclear matter. In the present analysis, we define the meson mass in nuclear matter as a pole of each propagator with vanishing three-momentum in which one-loop corrections by the nucleons in addition to the meson mean fields are included. A self energy including one loops in momentum space generally depends on the external momentum, but here we consider $\Pi_X(q_0, \vec{q}=\vec{0})$. In our approach, since the one loops are regarded as corrections to the mean-field approximation, we reduce the self-energy to a local form approximately as $\Pi_X(q_0,\vec{0}) \to \Pi_X(m_X,\vec{0})$ with $m_X$ a mass of meson $X$ in the mean-field level defined in Eqs.~(\ref{S1})-~(\ref{A1}). The concrete expressions of $\Pi_X(q_0,\vec{0})$ will be given in Appendix~\ref{sec:Spin0SE} and Appendix~\ref{sec:Spin1SE}. 

For convenience, let us define the quantity 
\begin{eqnarray}
\tilde{m}_X^2 \equiv m_X^2+\Pi_X(m_X,\vec{0})\ .
\end{eqnarray}
Then, as a naive extension of Eq.~(\ref{MassesNaive}), $a_0$, $\omega_N$, $\omega_S$, $\rho$, $K^*$, $f_{1N}$, $f_{1S}$, $a_1$, and $K_1$ masses in nuclear matter are easily provided by
\begin{eqnarray}
&&(m_{a_0}^2)^{\rm med} = \tilde{m}_{a_0}^2\, ,  \ (m_{\omega_N}^2)^{\rm med} = \tilde{m}_{\omega_N}^2\, , \ (m_{\omega_S}^2)^{\rm med} = \tilde{m}_{\omega_S}^2 \, , \nonumber\\
&& (m_\rho^2)^{\rm med} = \tilde{m}_\rho^2\, ,\ (m_{K^*}^2)^{\rm med} = \tilde{m}_{K^*}^2\, ,  \ (m_{f_{1N}}^2)^{\rm med}=\tilde{m}_{f_{1N}}^2\, , \nonumber\\
&& (m_{f_{1S}}^2)^{\rm med}=\tilde{m}_{f_{1S}}^2 \, , \ (m_{a_1}^2)^{\rm med}= \tilde{m}_{a_1}^2\, , \ (m_{K_1}^2)^{\rm med}= \tilde{m}_{K_1}^2\, . \nonumber\\  \label{Mass1}
\end{eqnarray} 
For the other mesons, we must solve the mixings as done in Appendix~\ref{sec:MeanMass}. Namely, as in Eq.~(\ref{MassesRenorm}) $\pi$, $K_0^*$, and $K$ masses in nuclear matter are 
\begin{eqnarray}
&&(m_\pi^2)^{\rm med} = {Z}_\pi^2\tilde{m}_\pi^2 \ , \ (m_{K_0^*}^2)^{\rm med} = {Z}_{K_0^*}^2\tilde{m}_{K_0^*}^2 \ , \nonumber\\
&&(m_K^2)^{\rm med} = {Z}_K^2\tilde{m}_K^2 \ ,  \label{Mass2}
\end{eqnarray}
with
\begin{eqnarray}
{Z}_{\pi} &=& \frac{\tilde{m}_{a_1}}{\sqrt{\tilde{m}_{a_1}^2-g_1^2{\phi}_N^2}} \ ,\nonumber\\
{Z}_{K_0^*} &=& \frac{2\tilde{m}_{K^*}}{\sqrt{4\tilde{m}_{K^*}^2-g_1^2({\phi}_N-\sqrt{2}{\phi}_S)^2}} \ ,\nonumber\\
{Z}_{K} &=& \frac{2\tilde{m}_{K_1}}{\sqrt{4\tilde{m}_{K_1}^2-g_1^2({\phi}_N+\sqrt{2}{\phi}_S)^2}} \ . \label{ZPiZK}
\end{eqnarray} 
For $\sigma_N$, $\sigma_S$, $\eta_N$, and $\eta_S$, we find
\begin{eqnarray}
&&({m}_{f_0^H/f_0^L}^2)^{\rm med} = \frac{1}{2}\Big(\tilde{m}_{\sigma_N}^2+\tilde{m}_{\sigma_S}^2 \nonumber\\
&&\ \ \ \ \ \ \ \ \ \ \ \  \ \ \ \ \ \ \ \  \pm\sqrt{(\tilde{m}_{\sigma_N}^2-\tilde{m}_{\sigma_S}^2)^2+4\tilde{m}_{\sigma_N\sigma_S}^4}\Big) \ , \nonumber\\
&& ({m}_{\eta'/\eta}^2)^{\rm med} = \frac{1}{2} \Big(({m}_{\eta_N}^2)^{\rm med}+({m}_{\eta_S}^2)^{\rm med}  \nonumber\\
&& \ \ \ \ \pm \sqrt{\big(({m}_{\eta_N}^2)^{\rm med}-({m}_{\eta_S}^2)^{\rm med} \big)^2+4({m}_{\eta_N\eta_S}^4)^{\rm med}}\Big)\ ,\nonumber\\ \label{Mass3}
\end{eqnarray}
where $(m_{\eta_N}^2)^{\rm med} = {Z}_{\eta_N}^2\tilde{m}_{\eta_N}^2$, $(m_{\eta_S}^2)^{\rm med} = {Z}_{\eta_S}^2\tilde{m}_{\eta_S}^2$ and $(m_{\eta_N\eta_S}^4)^{\rm med} = {Z}_{\eta_N}^2{Z}_{\eta_S}^2\tilde{m}_{\eta_N\eta_S}^4$ with
\begin{eqnarray}
{Z}_{\eta_N} &=& \frac{\tilde{m}_{f_{1N}}}{\sqrt{\tilde{m}_{f_{1N}}^2-g_1^2{\phi}_N^2}} \ ,\nonumber\\
{Z}_{\eta_S} &=& \frac{\tilde{m}_{f_{1S}}}{\sqrt{\tilde{m}^2_{f_{1S}}-2g_1^2{\phi}_S^2}} \ ,
\end{eqnarray}
as in Eq.~(\ref{MassF0Eta}) by introducing appropriate mixing angles.

\section{Self energies for the spin-$0$ mesons}
\label{sec:Spin0SE}
Here, we list explicit forms of self energies for the spin-$0$ mesons in nuclear matter with vanishing spatial momentum. The couplings with the nucleons for each meson can be read by the Lagrangian~(\ref{PDMThree}). By defining $E_k=\sqrt{|\vec{k}|^2+m_+^2}$, we can get the following results:\footnote{As stated in the main text, only the nucleon $N(939)$ forms a Fermi surface since we stick to lower density.}
\begin{widetext}
\begin{eqnarray}
{\Pi}_{\sigma_N}(q_0,\vec{0}) &=&  \frac{8\left(k_1\phi_N{\rm sin}\, 2\theta+g_{NN\sigma}\right)^2}{\pi^2}\int_0^{k_F} d|\vec{k}|\frac{|\vec{k}|^4}{E_k(4E_k^{2}-q_0^2)} + \frac{2m_+}{\pi^2}k_1{\rm sin}\, 2\theta\int_0^{k_F}d|\vec{k}|\frac{|\vec{k}|^2}{E_k} \nonumber\\
&&- \frac{4 \left(k_1\phi_N{\rm cos}\, 2\theta-g_{NN^*\sigma}\right)^2}{\pi^2}\int_0^{k_{F}} d|\vec{k}|\frac{|\vec{k}|^2}{E_k}\frac{2|\vec{k}|^2q_0^2+m_+(m_++m_-)(q_0^2-(m_+-m_-)^2)}{q_0^4-2(2|\vec{k}|^2+m_+^2+m_-^2)q_0^2+(m_+^2-m_-^2)^2} 
\ ,
\end{eqnarray}
\begin{eqnarray}
{\Pi}_{a_0}(q_0,\vec{0}) &=& \frac{8g_{NN\sigma}^2}{\pi^2}\int_0^{k_F} d|\vec{k}|\frac{|\vec{k}|^4}{E_k(4E_k^{2}-q_0^2)} -\frac{2m_+}{\pi^2}k_1{\rm sin}\, 2\theta\int_0^{k_F}d|\vec{k}|\frac{|\vec{k}|^2}{E_k}
 \nonumber\\
&&- \frac{4g_{NN^*\sigma}^2}{\pi^2}\int_0^{k_{F}} d|\vec{k}|\frac{|\vec{k}|^2}{E_k}\frac{2|\vec{k}|^2q_0^2+m_+(m_++m_-)(q_0^2-(m_+-m_-)^2)}{q_0^4-2(2|\vec{k}|^2+m_+^2+m_-^2)q_0^2+(m_+^2-m_-^2)^2} \ , \label{PiSigmaN}
\end{eqnarray}
\begin{eqnarray}
{\Pi}_{\eta_N}(q_0,\vec{0}) &=& \frac{8\left(k_2\phi_N{\rm sin}\, 2\theta+g_{NN\pi}\right)^2}{\pi^2}\int_0^{k_F} d|\vec{k}|\frac{|\vec{k}|^2E_k}{(4E_k^{2}-q_0^2)} -\frac{2m_+}{\pi^2}k_1 {\rm sin}\, 2\theta\int_0^{k_F}d|\vec{k}|\frac{|\vec{k}|^2}{E_k}\nonumber\\
&& - \frac{4\left(k_2\phi_N{\rm cos}\, 2\theta-g_{NN^*\pi}\right)^2}{\pi^2}\int_0^{k_{F}} d|\vec{k}|\frac{|\vec{k}|^2}{E_k}\frac{2|\vec{k}|^2q_0^2+m_+(m_+-m_-)(q_0^2-(m_++m_-)^2)}{q_0^4-2(2|\vec{k}|^2+m_+^2+m_-^2)q_0^2+(m_+^2-m_-^2)^2}\ , \label{PiEtaN}
\end{eqnarray}
and
\begin{eqnarray}
{\Pi}_{\pi}(q_0,\vec{0}) &=& \frac{8g_{NN\pi}^2}{\pi^2}\int_0^{k_F} d|\vec{k}|\frac{|\vec{k}|^2E_k}{(4E_k^{2}-q_0^2)} +\frac{2m_+}{\pi^2}k_1 {\rm sin}\, 2\theta\int_0^{k_F}d|\vec{k}|\frac{|\vec{k}|^2}{E_k} 
 \nonumber\\
&& - \frac{4g_{NN^*\pi}^2}{\pi^2}\int_0^{k_{F}} d|\vec{k}|\frac{|\vec{k}|^2}{E_k}\frac{2|\vec{k}|^2q_0^2+m_+(m_+-m_-)(q_0^2-(m_++m_-)^2)}{q_0^4-2(2|\vec{k}|^2+m_+^2+m_-^2)q_0^2+(m_+^2-m_-^2)^2} \ . \label{PiPi}
\end{eqnarray}
\end{widetext}
The remaining spin-$0$ mesons do not couple with the nucleons directly so that ${\Pi}_{\sigma_S}(q_0,\vec{0}) ={\Pi}_{K_0^*}(q_0,\vec{0}) = {\Pi}_{\eta_S}(q_0,\vec{0}) ={\Pi}_{K}(q_0,\vec{0}) = 0$.

\section{Self energies for the spin-$1$ mesons}
\label{sec:Spin1SE}
In this appendix, we show the explicit forms of self energies for spin-$1$ mesons in nuclear matter with vanishing spatial momentum. Before showing the results, to begin with, we discuss a general property of a spin-$1$ meson propagator in medium.

First, let us assume a propagator of a free spin-$1$ meson with mass $m$ in the vacuum takes the form of ``unitary gauge'':
\begin{eqnarray}
D_0^{\mu\nu}(q) = \frac{-i}{q^2-m^2}\left(g^{\mu\nu}-\frac{q^\mu q^\nu}{m^2}\right)\ ,
\end{eqnarray}
then, the inverse propagator is given by
\begin{eqnarray}
(D^{-1}_0)^{\mu\nu}(q) = i(q^2-m^2)g^{\mu\nu}-iq^\mu q^\nu
\end{eqnarray}
($q^2=q_0^2-|\vec{q}|^2$). Next, let us denote the self energy in medium by
\begin{eqnarray}
\Pi^{\mu\nu}(q_0,\vec{q}) &=& \Pi^T(q_0,\vec{q}) P_T^{\mu\nu}+\Pi^L(q_0,\vec{q})P_L^{\mu\nu} \nonumber\\
&& +\Pi^s(q_0,\vec{q})(g^{\mu\nu}-v^{\mu}v^\nu)+\Pi^t(q_0,\vec{q})v^\mu v^\nu\ ,\nonumber\\
\end{eqnarray}
where the three dimensional transverse and longitudinal projection operators are defined by
\begin{eqnarray}
P^T_{\mu\nu} = g_{\mu i}\left(\delta_{ij}-\frac{\vec{q}_i\vec{q}_j}{|\vec{q}|^2}\right)g_{j \nu}\ ,\ \ P^L_{\mu\nu} &=\frac{q_\mu q_\nu}{q^2}-g_{\mu\nu}-P^T_{\mu\nu}\ , \nonumber\\
\end{eqnarray}
and $v^\mu=(1,\vec{0})$ fixes the reference frame of the medium. Hence, according to the Dyson equation $(D^{-1})^{\mu\nu}(q_0,\vec{q}) = (D_0^{-1})^{\mu\nu}(q)-i\Pi^{\mu\nu}(q_0,\vec{q})$, we find the full propagator in medium expressed in terms of $\Pi^T(q_0,\vec{q})$, $\Pi^L(q_0,\vec{q})$, $\Pi^s(q_0,\vec{q})$, $\Pi^t(q_0,\vec{q})$ as
\begin{eqnarray}
D^{\mu\nu}(q_0,\vec{q}) &=&  \frac{i}{X^T(q_0,\vec{q})}P_T^{\mu\nu} +i\frac{q^2(m^2+\Pi^t(q_0,\vec{q}))}{X^L(q_0,\vec{q})}P_L^{\mu\nu}  \nonumber\\
&&+i\frac{q^2(q^2-m^2+\Pi^L(q_0,\vec{q})-\Pi^t(q_0,\vec{q}))}{X^L(q_0,\vec{q})}\frac{q^\mu q^\nu}{q^2} \nonumber\\
&& + i\frac{q^2(\Pi^t(q_0,\vec{q})-\Pi^s(q_0,\vec{q}))}{X^L(q_0,\vec{q})}v^\mu v^\nu \ , \label{MassiveSpin1}
\end{eqnarray}
with
\begin{eqnarray}
X^T(q_0,\vec{q}) &\equiv& q^2-m^2+\Pi^T(q_0,\vec{q})-\Pi^s(q_0,\vec{q}) \ ,\nonumber\\
X^L(q_0,\vec{q})&\equiv& q_0^2(m^2+\Pi^t(q_0,\vec{q}))(q^2-m^2+\Pi^L(q_0,\vec{q}) \nonumber\\
&&-\Pi^s(q_0,\vec{q}))-|\vec{q}|^2(m^2+\Pi^s(q_0,\vec{q}))(q^2-m^2 \nonumber\\
&& +\Pi^L(q)-\Pi^t(q))\ . \label{XTXL}
\end{eqnarray}
The transverse (longitudinal) mass of spin-$1$ meson is defined by the pole position of the full propagator in Eq.~(\ref{MassiveSpin1}) with vanishing spatial momentum: $X^{T(L)}(q_0,\vec{0})=0$. In calculating Eq.~(\ref{XTXL}), practically, it is useful to employ the following relations~\cite{Sasaki:2005yy}:
\begin{eqnarray}
\Pi^T(q_0,\vec{q})&=& \frac{1}{2}P_{T}^{\mu\nu}\Pi_{\mu\nu}+\frac{q_0q^i}{|\vec{q}|^2}\Pi^{i0}-\frac{q^iq^j}{|\vec{q}|^2}\Pi^{ij} \ ,\nonumber\\
\Pi^L(q_0,\vec{q})&=& \frac{q^2q^i}{q_0|\vec{q}|^2}\Pi^{i0} \ ,\nonumber\\
\Pi^s(q_0,\vec{q}) &=& \frac{q_0q^i}{|\vec{q}|^2}\Pi^{i0}-\frac{q^iq^j}{|\vec{q}|^2}\Pi^{ij} = \frac{q^i}{|\vec{q}|^2}q_\mu\Pi^{i\mu} \ ,\nonumber\\
\Pi^t(q_0,\vec{q}) &=& \Pi^{00}-\frac{q^i}{q_0}\Pi^{i0}  = \frac{1}{q_0}q_\mu\Pi^{\mu 0}\ . \label{abcd}
\end{eqnarray}

At first glance, $X^T(q_0,\vec{0})$ and $X^L(q_0,\vec{0})$ do not coincide by Eq.~(\ref{XTXL}), which allows us to define the two kinds of masses of spin-$1$ meson in medium. However, according to the explicit calculations in our model, we find $\Pi^T(q_0,\vec{0})=\Pi^L(q_0,\vec{0}) \equiv \Pi^V(q_0,\vec{0})$ and the full propopagator~(\ref{MassiveSpin1}) turns into
\begin{eqnarray}
D^{\mu\nu}(q_0,\vec{q}) &=& \frac{i}{q_0^2-m^2+\Pi^V(q_0,\vec{0})-\Pi^s(q_0,\vec{0})}P_T^{\mu\nu} \nonumber\\
&& +  \frac{i}{q_0^2-m^2+\Pi^V(q_0,\vec{0})-\Pi^s(q_0,\vec{0})} P_L^{\mu\nu} \nonumber\\
&&  + \cdots\ ,
\end{eqnarray}
which clearly shows that the masses of spin-$1$ meson in transverse and longitudinal components are identical as naively expected.

In the following we will show the results of self energies for spin-$1$ mesons in nuclear matter. The interaction terms are extracted by Eq.~(\ref{PDMThree}) as in Appendix~\ref{sec:Spin0SE}. We should note the self energy $\Pi_X(q_0,\vec{0})$ here is defined by $\Pi_X(q_0,\vec{0}) \equiv -\Pi_X^V(q_0,\vec{0})+\Pi^s_X(q_0,\vec{0})$. The results are
\begin{widetext}
\begin{eqnarray}
{\Pi}_{\omega_N}(q_0,\vec{0}) &=&   \frac{8}{3\pi^2} \left(\frac{1}{2}(g_V{\rm cos}^2\theta+h_V{\rm sin}^2\theta)+2\tilde{g}\right)^2\int_0^{k_F} d|\vec{k}|\frac{|\vec{k}|^2}{E_k}\frac{2|\vec{k}|^2+3m_+^2}{4E_k^{2}-q_0^2} \nonumber\\
&+& \frac{4}{3\pi^2}\left(\frac{1}{2}(g_V-h_V){\rm sin}\, \theta\, {\rm cos}\, \theta\right)^2\int_0^{k_{F}} d|\vec{k}|\frac{|\vec{k}|^2}{E_k} \frac{3m_+(m_+-m_-)\big((m_++m_-)^2-q_0^2\big)+2|\vec{k}|^2(m_+^2-m_-^2-2q_0^2)}{q_0^4-2(2|\vec{k}|^2+m_+^2+m_-^2)q_0^2+(m_+^2-m_-^2)^2}\ , \nonumber\\
\end{eqnarray}
\begin{eqnarray}
{\Pi}_{\rho}(q_0,\vec{0}) &=&   \frac{8}{3\pi^2} \left(\frac{1}{2}(g_V{\rm cos}^2\theta+h_V{\rm sin}^2\theta)\right)^2\int_0^{k_F} d|\vec{k}|\frac{|\vec{k}|^2}{E_k}\frac{2|\vec{k}|^2+3m_+^2}{4E_k^{2}-q_0^2} \nonumber\\
&+& \frac{4}{3\pi^2}\left(\frac{1}{2}(g_V-h_V){\rm sin}\, \theta\, {\rm cos}\, \theta\right)^2\int_0^{k_{F}} d|\vec{k}|\frac{|\vec{k}|^2}{E_k} \frac{3m_+(m_+-m_-)\big((m_++m_-)^2-q_0^2\big)+2|\vec{k}|^2(m_+^2-m_-^2-2q_0^2)}{q_0^4-2(2|\vec{k}|^2+m_+^2+m_-^2)q_0^2+(m_+^2-m_-^2)^2}\ , \nonumber\\
\end{eqnarray}
\begin{eqnarray}
{\Pi}_{f_{1N}}(q_0,\vec{0})&=&   \frac{16}{3\pi^2} \left(\frac{1}{2}(g_V{\rm cos}^2\theta-h_V{\rm sin}^2\theta)\right)^2\int_0^{k_F} d|\vec{k}|\frac{|\vec{k}|^2}{E_k}\frac{|\vec{k}|^2}{4E_k^{2}-q_0^2} \nonumber\\
&+& \frac{4}{3\pi^2}\left(\frac{1}{2}(g_V+h_V){\rm sin}\, \theta\, {\rm cos}\, \theta\right)^2\int_0^{k_{F}} d|\vec{k}|\frac{|\vec{k}|^2}{E_k} \frac{3m_+(m_++m_-)\big((m_+-m_-)^2-q_0^2\big)+2|\vec{k}|^2(m_+^2-m_-^2-2q_0^2)}{q_0^4-2(2|\vec{k}|^2+m_+^2+m_-^2)q_0^2+(m_+^2-m_-^2)^2}\ , \nonumber\\
\end{eqnarray}
and
\begin{eqnarray}
{\Pi}_{a_{1}}(q_0,\vec{0}) = {\Pi}_{f_{1N}}(q_0,\vec{0})\ .
\end{eqnarray}
\end{widetext}
The remaining spin-$1$ mesons do not couple with the nucleons directly so that ${\Pi}_{\omega_S}(q_0,\vec{0}) ={\Pi}_{K^*}(q_0,\vec{0}) = {\Pi}_{f_{1S}}(q_0,\vec{0}) ={\Pi}_{K_1}(q_0,\vec{0}) = 0$.

\end{document}